\documentclass[paper]{JHEP3}
\usepackage{epsfig}
\usepackage{amssymb}
\def    \be             {\begin{equation}}
\def    \ee             {\end{equation}}
\def    \ba             {\begin{eqnarray}}
\def    \ea             {\end{eqnarray}}

\def    \frac           #1#2{{#1 \over #2}}

\def    \bra#1          {\mbox{$\langle #1 |$}}
\def    \ket#1          {\mbox{$| #1 \rangle$}}

\newcommand\MCatNLO{MC@NLO}
\newcommand\pt{p_{\scriptscriptstyle \rm T}}
\newcommand\kt{k_{\scriptscriptstyle \rm T}}
\newcommand\sigtot{\sigma_{\rm tot}}

\def    \as             {\ifmmode\alpha_s\else$\alpha_s$\fi}
\def    \az             {\ifmmode\alpha_s^0\else$\alpha_s^0$\fi}

\def \sss{\scriptscriptstyle}
\def\mz{m_{\sss Z}}
\def\mzz{M_{\sss ZZ}}
\def\yzz{Y_{\sss ZZ}}

\newcommand\LambdaMSB{\Lambda_{\scriptscriptstyle \overline{\rm MS}}}

\newcommand\Nc{N_c}

\newcommand\qb{\overline{q}}

\newcommand\mqq{{\cal M}_{q\qb} }

\newcommand\fqq{f_{q\qb}}

\newcommand\fqg{f_{qg}}

\newcommand\thu{\theta_1}
\newcommand\thd{\theta_2}
\newcommand\omxplus{\left(\frac{1}{1-x}\right)_\rho}
\newcommand\omyplus{\left(\frac{1}{1-y}\right)_+}
\newcommand\opyplus{\left(\frac{1}{1+y}\right)_+}
\newcommand\lomxplus{\left(\frac{\log(1-x)}{1-x}\right)_\rho}

\newcommand\opmyplus{\left(\frac{1}{1\pm y}\right)_+}
\def\abs#1{\left|#1\right|} 
\newcommand\cf{C_{\sss\rm F}}
\newcommand\tf{T_{\sss\rm F}}

\newcommand\POWHEG{POWHEG}
\newcommand\nf{n_{\rm f}}
\newcommand\ptzz{p^{\rm \sss T}_{\rm \sss ZZ}}

\title{A Positive-Weight Next-to-Leading-Order Monte Carlo for
 $Z$ Pair Hadroproduction}
\vfill
\author{Paolo Nason \\
INFN, Sezione di Milano Bicocca, Italy\\
  E-mail: \email{Paolo.Nason@mib.infn.it}}
\author{Giovanni Ridolfi \\
Dipartimento di Fisica, Universit\`a di Genova\\
and INFN, Sezione di Genova, Italy\\
  E-mail: \email{Giovanni.Ridolfi@ge.infn.it}}
\abstract{
We present a first application of a previously published method for
the computation of QCD processes that is accurate at next-to-leading
order, and that can be
interfaced consistently to standard shower Monte Carlo programs.  We
have considered $Z$ pair production in hadron-hadron
collisions, a process whose complexity is sufficient to test
the general applicability of the method.
We have interfaced our result to the HERWIG and PYTHIA
shower Monte Carlo programs.  Previous work on next-to-leading order
corrections in a
shower Monte Carlo (the \MCatNLO{} program)
may involve the generation of events with negative weights,
that are avoided with the present method.
We have compared our results with those obtained with
\MCatNLO{}, and found remarkable consistency.
Our method can also be used as a standalone, alternative implementation
of QCD corrections, with the advantage of positivity, improved convergence,
and next-to-leading logarithmic accuracy in the region of small
transverse momentum of the radiated parton.
}
\preprint{Bicocca-FT-06-11 \\GEF-TH-1/2006\\
}

\begin{document}
\section{Introduction}

Next-to-leading order (NLO) QCD calculations have by now become the standard
for the study of processes relevant to collider physics. It is well
known that the inclusion of NLO terms are important for a reliable
estimate of cross sections. One finds corrections of the order of 30
to 100{\%} for typical production processes.  Until recently, the
results of NLO computations were not included in shower Monte Carlo
(SMC) models, since their inclusion is highly non trivial.  In
ref.~\cite{Frixione:2002ik} a method (referred to as \MCatNLO{}) was
proposed to include the results of NLO calculations in a parton shower
Monte Carlo, and it was applied to several processes\footnote{For the
full list of processes implemented in \MCatNLO, see the \MCatNLO{} web
page {\tt http://www.hep.phy.cam.ac.uk/theory/webber/MCatNLO/}.}
\cite{Frixione:2003ei,Frixione:2005vw} in connection with the HERWIG
implementation~\cite{Corcella:2002jc}.  The \MCatNLO{} method in
general must be adapted to the SMC algorithm being used. Furthermore,
the event generation is not strictly positive, i.e. negative weighted
events are generated.

In ref.~\cite{Nason:2004rx} a novel method was proposed, that overcomes
these problems, and allows one to compute the generation of the
hardest radiation (i.e. the radiation with the largest transverse
momentum) independently of the particular SMC
employed for the subsequent shower.  It was found there that,
in the framework of angular ordered SMC's,
in order
to preserve the correctness of the soft radiation spectrum, it is
necessary to include the soft radiation coherently emitted from the
partons produced in the hardest splitting.  Such coherent radiation
was named ``truncated shower'', since it is a shower that stops when
the angular variable approaches the angle of the hardest emission.

In ref.~\cite{Nason:2004rx}, only the general strategy was discussed,
while the path to follow for a practical implementation was not
studied. The
generation of the hardest event was only sketched there, and no
indication was given on how to actually perform it in practice.
Furthermore, the implementation of the truncated shower in a SMC program
was left to future work. This last problem should not however
be overemphasized. It is relevant only to angular ordered SMC's that
fully enforce soft coherence. At present, only the HERWIG program
does this. Other SMC's implementations do not care to enforce
coherence at the same level of accuracy.\footnote{
There are programs that generate soft radiation using dipole type formalisms
\cite{Gus88,Pet88}, but they do not treat initial state radiation
consistently.}
Furthermore, it may be possible to perform the generation of the truncated
shower independently of the particular SMC used.

 The problem of interfacing NLO calculations to SMC's can thus be split
into three independent issues: 
\begin{enumerate}
\item Generation of the hardest emission: one constructs an algorithm to
generate the hardest emission with NLO accuracy.
The resulting events can be put in a
standard form, such as the ``Les Houches
Interface for User Processes'' (LHI)~\cite{Boos:2001cv}.
\item Generation of the truncated shower:
given the hardest event,
on the LHI, one can add the truncated shower and put the event back
in the LHI. This step is only required if one wants to maintain
the correct soft emission pattern, and is using an angular ordered
SMC.
\item Showering, hadronization, decays: any SMC that complies with the
LHI requirements can now perform the rest of the showering, provided it
also implements a $\pt$ veto. Today's most popular SMC's satisfy
these requirements.
\end{enumerate}
Following the above strategy, the problem becomes more manageable,
since it no longer requires to modify or rewrite a full SMC
implementation.
Furthermore, several independent solutions to each of the above points may
be pursued by independent researchers.
In the present work, we tackle item 1 of the above list, in the
specific case of $Z$ boson pair production in hadronic collisions.
We have chosen this process for the following reasons:
\begin{itemize}
\item It is an important process for LHC physics.
\item It involves initial state radiation, which is more difficult
      than final state radiation.
\item It is readily extended to the similar $WZ$ and $WW$ production processes.
\item It should be easily generalizable to other important processes, like,
      for instance, single boson, Higgs and heavy flavour production.
\end{itemize}
We will now recall the basic formulae for the generation of the hardest emission
given in ref.~\cite{Nason:2004rx}, and illustrate in brief the
most relevant features of the method that we have developed.

In ref.~\cite{Nason:2004rx}, it was required that the phase space
kinematics is factorized in terms of Born ($v$) and radiation ($r$)
variables. In the case of $ZZ$ production the Born variables can be
chosen as the pair invariant mass $\mzz$ and rapidity $\yzz$, and the
cosine of a polar angle $\theta_1$.  In the case of two-body
subprocesses, $\theta_1$ is the angle formed by the direction of one
of the outgoing $Z$'s and the beam axis, in the rest frame of the $ZZ$
system. In the case of ${\cal O}(\as)$ real-emission corrections,
the $ZZ$ system has non-zero transverse momentum.
In this case we define a three-dimensional frame in the
$ZZ$ rest system, and define
$\theta_1$ as the angle of one of the $Z$'s relative to its third axis.
There is some arbitrariness in the choice of this frame.
The only important requirement is that in the limit
of zero transverse momentum of the $ZZ$ pair, its third axis
should become parallel to the
collision axis.
A specific choice is described in detail in ref.~\cite{Mele:1990bq}.
 The radiation variables $(r)$
are chosen to be $x=\mzz^2/s$, where $s$ is the invariant mass of the
incoming partonic system, a variable $-1<y<1$, equal to the cosine of
the scattering angle of the emitted parton in the rest frame of the
incoming parton system, and an azimuthal variable $\theta_2$ which
parametrizes the relative azimuthal angle of the $ZZ$ system with
respect to the radiated parton. With these definitions, when the $r$
variables approach the soft ($x\to 1$) or collinear ($y\to \pm
1$) limits, the kinematics of the $ZZ$ system approaches the
corresponding Born kinematics with the given $v$ variables.
We denote by
\be 
B(v)d\Phi_v,\quad V(v) d\Phi_v,\quad R(v,r)\, d\Phi_v\, d\Phi_r, \quad
C(v,r)\, d\Phi_v\, d\Phi_r
\ee
the Born, soft-virtual, real and
counterterm contributions to the cross section, respectively.  The differential
cross section for the hardest emission can be written schematically as
\begin{equation} \label{eq:introsigmahard}
d\sigma = {\bar B}(v) d\Phi_v
\left[\Delta(v,0)+\Delta(v,\kt(v,r))\frac{R(v,r)}{B(v)}d\Phi_r \right]\,,
\end{equation}
where
\ba \label{eq:introbbar}
{\bar B}(v)&=&B(v)+V(v)+\int d\Phi_r \,
\left[ R(v,r)-C(v,r)\right]
\\ \label{eq:introdelta}
\Delta(v,\pt)&=&\exp \left[ - \int
\frac{R(v,r)}{B(v)} \theta(\kt(v,r)-\pt)\,d\Phi_r\right]\,,
\ea
and $\kt(v,r)$ is the transverse momentum of the emitted parton.
As written above the calculation of the probability of the hardest
emission would seem computationally intensive. In particular,
the computation of ${\bar B}(v)$ requires one three-dimensional
integral for each $v$. In order to circumvent this problem, we introduce
the function
\begin{equation}
{\tilde B}(v,r) = N[B(v)+V(v)]+R(v,r)-C(v,r)
\end{equation}
where
\begin{equation}
N=\frac{1}{\int d\Phi_r}\;.
\end{equation}
so that
\begin{equation}
{\bar B}(v)=\int {\tilde B}(v,r)d\Phi_r\;.
\end{equation} 
Standard procedures are available to generate unweighted $v,r$ events
from the distribution ${\tilde B}(v,r)d\Phi_r\,d\Phi_v\;$.
This is exactly what we need: we generate $v,r$ values in this way,
and we ignore the $r$ value, which amounts to integrating over the
$r$ variables. Thus, the generation of unweighted events distributed
according to ${\bar B}(v)$ is no more expensive, from a computational
point of view, than generating unweighted events for the real emission
matrix elements.
The generation of the radiation variables $r$ also looks computationally
intensive, but it can be performed in an efficient way by using
the veto method.

In the following, we will illustrate all the details of the implementation
of the procedure outlined above.
In Section~\ref{sec:kin} we collect the kinematics and cross section formulae
for the $ZZ$ production process. In Section~\ref{sec:hard} we write down in
full detail eq.~(\ref{eq:introsigmahard}). In Section~\ref{sec:scales}
we discuss important issues having to do with the choice of scales,
and the accuracy in the Sudakov form factors for the generation of the
hardest event.
In Section~\ref{sec:hardgen} we first describe how one would perform
a straightforward implementation of unweighted event generation using
the given hard cross section. In Subsections \ref{sec:borngen} and
\ref{sec:radgen} we illustrate in detail our method for the generation
of the Born variables $v$ and of the radiation variables $r$.
Explanations of the Monte Carlo techniques that we have used are
reported in detail in the appendices, for ease of reference.

\section{Kinematics and cross section}\label{sec:kin}

The differential cross section for the production of $Z$ boson pairs
in hadronic collisions was computed in ref.~\cite{Mele:1990bq} up to
order $\as$. The effects of spin correlations of the decay
products of the two vector bosons have been computed in 
refs.~\cite{Dixon:1999di,Campbell:1999ah}, but will not
be included here for simplicity.
In this section, we formulate the result of
ref.~\cite{Mele:1990bq} in a form which is suitable for the
generation of the hardest event using the
procedure proposed in ref.~\cite{Nason:2004rx}.

The order-$\as$ cross section for the process $H_1 H_2\rightarrow
ZZ+X$ can be written as the sum of four terms:
\be
d\sigma=d\sigma^{\rm (b)}+d\sigma^{\rm (sv)}+d\sigma^{\rm (f)}
+d\sigma^{\rm (c)}.
\ee
Here $d\sigma^{\rm (b)}$ is the leading-order (Born) cross section.  The
term $d\sigma^{\rm (sv)}$ collects order-$\as$ contributions with the same
two-body kinematics as the Born term, namely one-loop corrections
and real-emission contributions in the soft limit. Finally,
$d\sigma^{\rm (f)}$ represents the cross section for real emission in a
generic configuration, while $d\sigma^{\rm (c)}$ is a remnant of the
subtraction of collinear singularities, and describes
real emission in the collinear limit.

At leading order, the only relevant parton subprocess is
\be
q(p_1)+\bar q(p_2)\rightarrow Z(k_1)+ Z(k_2),
\label{twobody}
\ee
where $q$ is a quark or antiquark of any flavour,
and $\bar q$ the corresponding antiparticle.
Particle four-momenta are displayed in brackets;
we have $p_1^2=p_2^2=0$, $k_1^2=k_2^2=\mz^2$, where $\mz$ is the
$Z$ boson mass. We introduce the usual Mandelstam
invariants
\be
s=(p_1+p_2)^2,\qquad
t=(p_1-k_1)^2,\qquad
u=(p_1-k_2)^2,
\ee
related by $s+t+u=2\mz^2$.

Event generation is conveniently performed in terms of the invariant mass
$\mzz$ and the rapidity $\yzz$ of the $Z$ boson pair
in the laboratory frame. They are given by
\ba
\label{M2xx}
&&\mzz^2=(k_1+k_2)^2=x_1\,x_2\,S
\\
\label{Yxx}
&&\yzz=\frac{1}{2}\log\frac{(p_1+p_2)^0+(p_1+p_2)^3}
{(p_1+p_2)^0-(p_1+p_2)^3}=\frac{1}{2}\log\frac{x_1}{x_2},
\ea
where $S$ is the squared center-of-mass total energy
of the colliding hadrons, $x_1$ and $x_2$ are the fractions of longitudinal
momenta carried by the incoming partons, and we have used the fact that
the $ZZ$ pair has zero rapidity in the 
center-of-mass frame of the colliding partons.
It follows that
\be
x_1=\sqrt{\frac{\mzz^2}{S}}\,e^{\yzz}\equiv x_{b1};\quad
x_2=\sqrt{\frac{\mzz^2}{S}}\,e^{-\yzz}\equiv x_{b2};\quad
dx_1 dx_2=\frac{1}{S}\,d\yzz d\mzz.
\label{bornjac}
\ee
We adopt as two-body kinematic variables the set
$v=\{\mzz,\yzz,\cos\theta_1\}$ (which we will call the Born variables
henceforth), where $\theta_1$ is the angle between
$\vec{p}_1$ and $\vec{k}_1$ in the partonic center-of-mass frame,
so that
\be
t=\mz^2-\frac{\mzz^2}{2}(1-\beta\cos\theta_1)\,,
\label{tborn}
\ee
where
\be
\beta=\sqrt{1-\rho};\qquad \rho=\frac{4\mz^2}{\mzz^2}\,.
\label{betadef}
\ee
Using eq.~(\ref{bornjac}) and the usual expression
of the two-body phase space measure $d\Phi_2$, 
it is immediate to check that
\be
d\Phi_2\,dx_1\,dx_2=
\frac{\beta}{16\pi S}\,d\cos\theta_1\,d\mzz^2\,d\yzz\,.
\ee
In order to keep our notation similar to that of
ref.~\cite{Nason:2004rx}, we define
\be
\label{phiv}
d\Phi_v=d\cos\theta_1\,d\mzz^2\,d\yzz.
\ee
The appropriate integration region for the variables $v$ is
\be
4\mz^2\leq \mzz^2 \leq S\,,\quad
\frac{1}{2}\log\frac{\mzz^2}{S}\leq \yzz\leq
-\frac{1}{2}\log\frac{\mzz^2}{S}\,,\quad
-1\leq\cos\theta_1\leq 1.
\ee

The Born cross section is given by
\be
d\sigma^{\rm (b)}=d\Phi_v\,\sum_q B_q(v,\mu),
\ee
where
\be
\label{eq:bqdef}
B_q(v,\mu)=
\frac{\beta}{16\pi S}\,
f_q^{H_1}(x_{b1},\mu)\, f_{\bar q}^{H_2}(x_{b2},\mu)\,
{\cal M}_{q\bar q}^{\rm (b)}(\mzz^2,t).
\ee
The index $q$ runs over all quarks and antiquarks; $f_q^H(x,\mu)$ denotes
the distribution function of parton $q$ in the hadron $H$, and
$\mu$ is a factorization scale. The function 
${\cal M}_{q\bar q}^{\rm (b)}(s,t)$ is the squared invariant amplitude,
summed over final-state polarizations and averaged over initial-state
polarizations and colors, divided by the relevant flux factor:
\be
{\cal M}_{q\bar q}^{\rm (b)}(s,t)=
\frac{1}{2s} \frac{g_{q,V}^4+g_{q,A}^4+6 g_{q,A}^2 g_{q,V}^2}{N_C}
\,\left[\frac{t}{u}+\frac{u}{t}+\frac{4\mz^2s}{tu}
-\mz^4\left(\frac{1}{t^2}+\frac{1}{u^2}\right)\right]
\label{Born}
\ee
where $g_{q,V}$ and $g_{q,A}$ denote the vector
and axial-vector couplings of the quark $q$ to the $Z$ boson,
and $N_C=3$ is the number of colours. 

Order-$\as$ contributions to the cross section arise from one-loop
corrections to the two-body process eq.~(\ref{twobody}), and from
real-emission subprocesses at tree level. The
contribution of one-loop diagrams must be summed to the one-gluon
emission cross section in the soft limit, in order to obtain an
infrared-finite result.  The resulting contribution has the same
kinematic structure as the leading-order term:
\be
d\sigma^{\rm (sv)}=d\Phi_v\sum_q V_q(v,\mu),
\ee
where
\ba
V_q(v,\mu)&=&\frac{\beta}{16\pi S}\,
f_q^{H_1}(x_{b1},\mu)\, f_{\bar q}^{H_2}(x_{b2},\mu)
\Bigg\{\mqq^{(v,{\rm reg})}(\mzz^2,t)
\label{softvirt}\\
&& +\frac{C_F\as(\mu^2)}{4\pi}
\left[\log\frac{\mzz^2}{\mu^2}(6+16\log\beta)+32\log^2\beta
-\frac{4}{3}\pi^2\right] \mqq^{\rm (b)}(\mzz^2,t,\mu)\Bigg\}, \nonumber
\ea
and $C_F=4/3$. The explicit expression of $\mqq^{(v,{\rm
reg})}(s,t,\mu)$ is given in Appendix B of ref.~\cite{Mele:1990bq}. The $\mu$
dependence in $\mqq^{(v,{\rm reg})}$ is implicit through its
dependence upon $\as$. Notice also that the explicit $\mu$ dependence
in eq.~(\ref{softvirt}) is due to the collinear subtraction.  For
simplicity, we have chosen the same value for the renormalization and
factorization scales.  Since the Born process is of order $0$ in
$\as$, there is no explicit $\mu$ dependence due to renormalization,
and therefore the renormalization scale only appears as the argument
of $\as$.

Next, we consider the real-emission subprocesses
\ba
q(p_1)+\bar q(p_2)&\rightarrow& Z(k_1)+Z(k_2)+g(k)
\label{qq}
\\
q(p_1)+g(p_2)&\rightarrow& Z(k_1)+Z(k_2)+q(k)
\label{qg}
\\
g(p_1)+\bar q(p_2)&\rightarrow& Z(k_1)+Z(k_2)+\bar q(k)
\label{gq}
\ea
in a generic kinematical configuration.
The processes (\ref{qq}-\ref{gq})
are characterized by five independent scalar quantities, which
we choose to be
\be
s=(p_1+p_2)^2,\quad
t_k=(p_1-k)^2,\quad
u_k=(p_2-k)^2, \quad
q_1=(p_1-k_1)^2,\quad
q_2=(p_2-k_2)^2,
\ee
as in ref.~\cite{Mele:1990bq}. We introduce the variables 
\be
x=\frac{\mzz^2}{s};\qquad y=\cos\theta,
\ee
where $\theta$ is the scattering angle of the emitted parton
in the partonic center-of-mass system. With these definitions,
\be
t_k=-\frac{s}{2}(1-x)(1-y);\qquad
u_k=-\frac{s}{2}(1-x)(1+y).
\label{tandu}
\ee
It is easy to show that in the case of the subprocesses
(\ref{qq}-\ref{gq}) one has
\be
\yzz=\frac{1}{2}\log\frac{x_1}{x_2}\frac{s+u_k}{s+t_k}
=\frac{1}{2}\log\frac{x_1}{x_2}\frac{2-(1-x)(1+y)}{2-(1-x)(1-y)};
\qquad
\mzz^2=x\,x_1x_2\,S,
\ee
and therefore
\be
x_1=\frac{x_{b1}}{\sqrt{x}}\,\sqrt{\frac{2-(1-x)(1-y)}{2-(1-x)(1+y)}};
\qquad
x_2=\frac{x_{b2}}{\sqrt{x}}\,\sqrt{\frac{2-(1-x)(1+y)}{2-(1-x)(1-y)}}
\label{x1x2}
\ee
and
\be
\label{mcjac}
dx_1\,dx_2=\frac{1}{xS}\,d\mzz^2\,d\yzz.
\ee
The range for the variable $x$ is restricted by the requirement
that both $x_1$ and $x_2$ be less than 1; this gives
\be
x_{\rm min}\leq x\leq 1,
\ee
with
\ba
x_{\rm min}&=& {\rm max}\left(
\frac{2(1+y)\,x_{b1}^2}{\sqrt{(1+x_{b1}^2)^2(1-y)^2+16yx_{b1}^2}
    +(1-y)(1-x_{b1}^2)},\right.
\nonumber\\
&&\phantom{aaaaa}\left.
\frac{2(1-y)\,x_{b2}^2}{\sqrt{(1+x_{b2}^2)^2(1+y)^2-16yx_{b2}^2}
    +(1+y)(1-x_{b2}^2)}\right).
\label{xmin}
\ea
Note that $x_{\rm min}$ depends explicitly on $y$,
and implicitly on $\mzz^2$ and $\yzz$ through $x_{b1},x_{b2}$.
It can be checked that $x_{\rm min}$ is always larger that $\mzz^2/S$,
as required by the definition of~$x$.

In the center-of-mass frame of the $ZZ$ system,
the four-momenta of the produced $Z$ bosons can be parametrized
in terms of two angles $\theta_1,\theta_2$:
\ba
k_1&=&\frac{\mzz}{2}\;
(1,\beta\sin\theta_2\sin\theta_1,
\beta\cos\theta_2\sin\theta_1,\beta\cos\theta_1) \nonumber \\
k_2&=&\frac{\mzz}{2}\;
(1,-\beta\sin\theta_2\sin\theta_1,
-\beta\cos\theta_2\sin\theta_1,-\beta\cos\theta_1),
\ea
with $\beta$ given in eq.~(\ref{betadef}).
Both $\theta_1$ and $\theta_2$ range between $0$ and $\pi$.
Thus, in addition to the Born variables $v$, 
we have now the three radiation variables $r=\{x,y,\thd\}$, with
\be
x_{\rm min}\leq x\leq 1\,,\quad
-1\leq y\leq 1\,,\quad
0\leq\thd\leq\pi\,.
\ee
Following ref.~\cite{Nason:2004rx}
we define the corresponding integration measure
\be
\label{dphir}
d\Phi_r=dx\,dy\,d\thd.
\ee

From the computation of ref.~\cite{Mele:1990bq} we obtain
\be
d\sigma^{\rm (f)}=d\Phi_v\,d\Phi_r\,
\sum_q \left[R_{q\bar q}(v,r,\mu)+R_{qg}(v,r,\mu)
+R_{g\bar q}(v,r,\mu)\right],
\ee
where
\ba
R_{q\bar q}(v,r,\mu)&=&
\frac{1}{(4\pi)^2}
\frac{\beta}{64\pi^2 \mzz^2 S}\,
\omxplus\left[\omyplus+\opyplus\right]
\label{eq:rdef}
\\
&&f_q^{H_1}(x_1,\mu)\, f_{\bar q}^{H_2}(x_2,\mu)\,
\fqq(x,y,\thu,\thd,\mu)
\nonumber\\
R_{qg}(v,r,\mu)&=&
\frac{1}{(4\pi)^2}
\frac{\beta}{32\pi^2 \mzz^2 S}
\opyplus f_q^{H_1}(x_1,\mu)\, f_g^{H_2}(x_2,\mu)\,
\fqg(x,y,\thu,\thd,\mu)
\nonumber\\
R_{g\bar q}(v,r,\mu)&=&
\frac{1}{(4\pi)^2}
\frac{\beta}{32\pi^2 \mzz^2 S}
\omyplus f_g^{H_1}(x_1,\mu)\, f_{\bar q}^{H_2}(x_2,\mu)\,
\fqg(x,-y,\thu,\thd,\mu).
\nonumber
\ea
The functions 
$R_{q\bar q}$, $R_{qg}$, $R_{g\bar q}$
denote the regularized
real emission cross sections for the different subprocesses.
The functions $\fqq$ and $\fqg$ are regular in the limits
of soft ($x=1$) or collinear ($y=\pm 1$) emission;
they are given explicitly in eqs.~(2.26,2.66) and Appendix~C
of ref.~\cite{Mele:1990bq}.
The distributions $1/(1-x)_\rho$ and $1/(1\pm y)_+$
are defined by
\ba
\label{omxdef}
&&\int_\rho^1dx\,g(x)\,\omxplus=
\int_\rho^1dx\,\frac{g(x)-g(1)}{1-x}
\\
&&\int_{-1}^1dy\,h(y)\,\opmyplus=
\int_{-1}^1dy\,\frac{h(y)-h(\mp 1)}{1\pm y}.
\label{opmydef}
\ea

As shown in ref.~\cite{Mele:1990bq}, the remnants
of the collinear subtraction must also be added to get the
full cross section. This contribution has the form
\ba
d\sigma^{\rm (c)}&=&d\Phi_v\,dx\,dy\,\sum_q\Big[
\left( L^+_{q\bar q}(v,x,\mu)+L_{g\bar q}(v,x,\mu) \right) \,\delta(1-y)
\nonumber \\
&& +\left(L^-_{q\bar q}(v,x,\mu)
+L_{qg}(v,x,\mu)\right) \,\delta(1+y)\Big]\,,
\ea
where
\ba
L^+_{q\bar q}(v,x,\mu)&=&
\frac{C_F\as}{2\pi}\frac{\beta}{16\pi S}
\left\{\left[\log\frac{\mzz^2}{x\mu^2}\omxplus
+2\lomxplus\right](1+x^2)+(1-x)\right\}
\nonumber\\  \label{eq:lplus}
&\times&
\mqq^{\rm (b)}(\mzz^2,t)\,
f_q^{H_1}(x_{b1}/x,\mu)\,f_{\bar q}^{H_2}(x_{b2},\mu)
\\
L^-_{q\bar q}(v,x,\mu)&=&
\frac{C_F\as}{2\pi}\frac{\beta}{16\pi S}
\left\{\left[\log\frac{\mzz^2}{x\mu^2}\omxplus
+2\lomxplus\right](1+x^2)+(1-x)\right\}
\nonumber\\ \label{eq:lminus}
&\times&
\mqq^{\rm (b)}(\mzz^2,t)\,
f_q^{H_1}(x_{b1},\mu)\,f_{\bar q}^{H_2}(x_{b2}/x,\mu)
\\
L_{g\bar q}(v,x,\mu)&=&
\frac{T_F\as}{2\pi}
\frac{\beta}{16\pi S}
\left[\left(\log\frac{\mzz^2}{x\mu^2}+2\log(1-x)\right)(x^2+(1-x)^2)
+2x(1-x)\right] 
\nonumber\\ \label{eq:colgq}
&\times&
\mqq^{\rm (b)}(\mzz^2,t)\,
f_g^{H_1}(x_{b1}/x,\mu)\, f_{\bar q}^{H_2}(x_{b2},\mu)
\\
L_{qg}(v,x,\mu)&=&
\frac{T_F\as}{2\pi}
\frac{\beta}{16\pi S}
\left[\left(\log\frac{\mzz^2}{x\mu^2}+2\log(1-x)\right)(x^2+(1-x)^2)
+2x(1-x)\right] 
\nonumber\\ \label{eq:colqg}
&\times&
\mqq^{\rm (b)}(\mzz^2,t)\,
f_q^{H_1}(x_{b1},\mu)\, f_g^{H_2}(x_{b2}/x,\mu),
\ea
and $t$ is given in eq.~(\ref{tborn}).
From eq.~(\ref{xmin}) we see that
the integration range becomes
$x_{b1}<x<1$ for eqs.~(\ref{eq:lplus},\ref{eq:colgq})
and $x_{b2}<x<1$ for eqs.~(\ref{eq:lminus},\ref{eq:colqg}).

\section{Hardest event cross section}\label{sec:hard}
\begin{figure}[ht]
\begin{center}
\epsfig{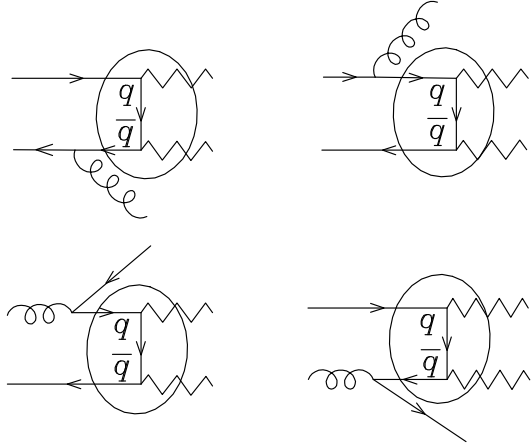}
\end{center}
\caption{\label{fig:lump}
NLO subprocesses that share the same elementary flavour
structure.}
\end{figure}

In this section, we will write eq.~(\ref{eq:introsigmahard})
for the case at hand in full detail.
The method of ref.~\cite{Nason:2004rx}, when applied to a generic
process, may require a separated treatment of each singular region.
In this case this is not needed.
Our choice of variables $v,r$ is adequate for both collinear
regions at the same time, the only difference being the sign of $y$.
We have instead to pay attention to the
flavour structure of the process.
In ordinary SMC codes, the flavour structure of the Born
subprocess is not altered by subsequent radiation.
On the other hand, if
the hardest radiation is produced in the context of a NLO
calculation, the association of the NLO process to
a Born subprocess is not always obvious. A given real-emission
contribution must be associated to the Born
process in which it factorizes in the collinear limit.
In the present case, the collinear regions for the $q\bar{q}$
subprocess always factorize in terms of the $q\bar{q}$
Born process, and the same holds for the collinear regions of the
$g\bar q$ and $qg$ subprocesses, as shown in fig.~\ref{fig:lump}.
Thus, for a given flavour~$q$, we lump together the
$q\bar{q}$, the $qg$ and the $g\bar{q}$ real-emission subprocesses.

Following ref.~\cite{Nason:2004rx}, we write the cross section for
the hardest event as
\begin{equation}\label{eq:hardestsigma}
d\sigma = \sum_q {\bar B}_q(v,\mu_v) d\Phi_v \left[
\Delta_q(0)+\Delta_q(\kt)
\frac{{\hat R}_{q\bar{q}}(v,r,\mu_r)+{\hat R}_{qg}(v,r,\mu_r)
+{\hat R}_{g\bar{q}}(v,r,\mu_r)}{B_q(v,\mu_r)}
d\Phi_r \right]\,,
\end{equation}
where the $\hat{\phantom{R}}$ on top of $R$ means that we drop
the $+$ prescriptions that regularize the $x$ and $y$ singularities.
The $\hat{R}$ are thus the unregularized real emission cross sections
(corresponding to $R$ in the notation of
ref.~\cite{Nason:2004rx}).
Furthermore,
\ba
{\bar B}_q(v,\mu) &=& 
B_q(v,\mu)+V_q(v,\mu)+\int d\Phi_r \,
\left[ R_{q\bar{q}}(v,r,\mu)+R_{qg}(v,r,\mu)
+R_{g\bar{q}}(v,r,\mu)\right]
\nonumber \\
&&+\int_{-1}^1 dy \int_{x_{\rm min}}^1 dx\,
\left[L^+_{q\bar{q}}(v,x,\mu)+L_{g\bar{q}}(v,x,\mu)\right]\,\delta(1-y)
\nonumber \\ \label{eq:bbdef}
&&+\int_{-1}^1 dy \int_{x_{\rm min}}^1 dx\,
\left[L^-_{q\bar{q}}(v,x,\mu)+L_{qg}(v,x,\mu)\right]\,\delta(1+y),
\ea
\be
\label{eq:Deltaq}
\Delta_q(\pt)=\exp \left[ - \int
\frac{{\hat R}_{q\bar{q}}(v,r,\mu_r)+{\hat R}_{qg}(v,r,\mu_r)
+{\hat R}_{g\bar{q}}(v,r,\mu_r)}{B_q(v,\mu_r)} \theta(\kt(v,r)-\pt)
 d\Phi_r\right]\,,
\ee
and $\kt(v,r)$ is the transverse momentum of the radiated parton,
\begin{equation} \label{eq:ktdef}
\kt(v,r)=\sqrt{\frac{\mzz^2}{4x}(1-x)^2\,(1-y^2)}\;.
\end{equation}
Equation~(\ref{eq:hardestsigma}) is the analogue of eq.~(5.10)
of ref.~\cite{Nason:2004rx}. The function $\Delta_q(\pt)$ corresponds to
$\Delta_R^{(\rm NLO)}(\pt)$ in the notation of ref.~\cite{Nason:2004rx}.

\section{Scale choices and Sudakov form factor}\label{sec:scales}
The scale choices in eqs.~(\ref{eq:hardestsigma}-\ref{eq:Deltaq})
deserve particular
attention. We denote by $\mu_v$ a scale that depends
only upon the Born variables,
and by $\mu_r$ a scale that is appropriate
to the radiation process, that is to say, of order
$\kt$. Appropriate choices are, for example, $\mu_v=\mzz$ and
$\mu_r=\kt(v,r)$. Thus, the distinction between these two scales becomes
particularly important when $\kt(v,r)\ll \mzz$, which is in fact the
region where most radiation is produced. The correct scale choice
is important here to maintain leading log (LL) accuracy
in the Sudakov form factor, eq.~(\ref{eq:Deltaq}).

We now remind the reader how the log counting is done in the Sudakov exponent.
We call $L$ the large logarithm
$\log(Q^2/\pt^2)$, where $Q$ is a scale of the order of the upper limit
for $\kt$. The dominant terms in the exponent have the structure
$L(\as(Q^2) L)^k$, $k=1,\ldots,\infty$. We call these
the LL terms. The NLL terms are of order $(\as L)^k$, the
NNLL terms $\as(\as L)^k$, and so on.
Roughly speaking, in the Sudakov exponent, the $dx/(1-x)$
singularity and the $dy/(1-y^2)$ singularity both contribute a
factor of $L$. Expanding $\as(\kt)$ as
\begin{equation}
\as(\kt^2)=\as(Q^2)\left[1+\sum_{k=1}^\infty \left(b_0\as(Q^2)
\log\frac{Q^2}{\kt^2}\right)^k\right]
\end{equation}
we see that LL terms are generated when both the $x$ and $y$ singularities are
present, and NLL terms are generated when only one of them is present.
Terms with no singularities are of NNLL importance.

We now show how, by a suitable slight modification
of $\as$, we can achieve
next-to-leading (NLL) accuracy in the Sudakov form factor.
To see this, we isolate the singular part in the integrand
of the Sudakov exponent
\newcommand\onetwo{\bigg\{{
\begin{array}{c}
\scriptstyle
 1\leftrightarrow 2 \\[-6pt]
\scriptstyle
 q \leftrightarrow \bar{q} \\[-6pt]
\scriptstyle
 y \leftrightarrow -y
\end{array}}\bigg\}}
\begin{eqnarray}
&& \frac{{\hat R}_{q\bar{q}}(v,r,\mu_r)+{\hat R}_{qg}(v,r,\mu_r)
+{\hat R}_{g\bar{q}}(v,r,\mu_r)}{B_q(v,\mu_r)}
=
\frac{\cf\as}{2\pi} \frac{1}{\pi}
\frac{1+x^2}{(1-y)(1-x)}
\frac{f^{H_1}_q(x_{b1}/x,\mu_r)}{ x\,f^{H_1}_q(x_{b1},\mu_r)}
\nonumber \\&& \label{eq:singlimit}
+\frac{\tf\as}{2\pi} \frac{1}{\pi}\frac{x^2+(1-x^2)}{1-y}
\frac{f^{H_1}_g(x_{b1}/x,\mu_r)}{x\, f^{H_1}_q(x_{b1},\mu_r)}
 + \onetwo
+\mbox{regular terms}\;.
\end{eqnarray}
The above equation is a consequence of collinear and soft factorization,
and can be easily obtained from eqs.~(2.26), (2.42), (2.61) and (2.66) of
ref.~\cite{Mele:1990bq}, together with the definitions of $B$ and $R$,
eqs.~(\ref{eq:bqdef}) and (\ref{eq:rdef}) in this paper.
We now replace
\begin{equation}
\frac{f_q^{H_{1}}(x_{b1}/x,\mu_r)}{xf_q^{H_{1}}(x_{b1},\mu_r)}=
\left[\frac{f_q^{H_{1}}(x_{b1}/x,\mu_r)}{xf_q^{H_{1}}(x_{b1},\mu_r)}-1\right]+1
\end{equation}
so that the term in square bracket vanishes for $x\to 1$.
Then we perform a change of variable, trading $y$ for $\kt$,
according to the formula
\begin{equation}
\frac{d\kt^2}{\kt^2} = \frac{2y dy}{1-y^2}\,,\qquad y=\pm\sqrt{1-\frac{4x\kt^2}{(1-x)^2\mzz^2}}\,,
\end{equation}
and work out the Sudakov exponent up to NLL accuracy.
We obtain
\begin{eqnarray}
\log \Delta_q(\pt)&\simeq& 
-\int_{\pt^2}^{Q^2} \frac{d\kt^2}{\kt^2} \frac{\cf\as(\kt^2)}{2\pi}
\int_0^1 dx\, \frac{1+x^2}{1-x} \left[\frac{f_q^{H_{1}}(x_{b1}/x,\kt)}{xf_q^{H_{1}}(x_{b1},\kt)}-1\right]
+\onetwo
\nonumber \\&&
-\int_{\pt^2}^{Q^2} \frac{d\kt^2}{\kt^2}\frac{\tf\as(\kt^2)}{2\pi} \int_0^1 dx\, [x^2+(1-x)^2] \,
\frac{f_g^{H_{1}}(x_{b1}/x,\kt)}{x f_q^{H_{1}}(x_{b1},\kt)}
+\onetwo
\nonumber \\&&
-\int_{\pt^2}^{Q^2} \frac{d\kt^2}{\kt^2}\frac{\cf\as(\kt^2)}{\pi} 
\int_0^{1-\frac{2\kt}{\mzz}} \frac{ dx}{\sqrt{1-\frac{4\kt^2}{(1-x)^2\mzz^2}}} \frac{1+x^2}{1-x}
\nonumber \\
&\simeq & -\int_{\pt^2}^{Q^2} \frac{d\kt^2}{\kt^2} \frac{d\log f_q^{H_{1}}(x_{b1},\kt)}{d\log\kt^2}
 -\int_{\pt^2}^{Q^2} \frac{d\kt^2}{\kt^2} \frac{d\log f_{\bar q}^{H_{2}}(x_{b2},\kt)}{d\log\kt^2}
\nonumber \\&&
-\int_{\pt^2}^{Q^2} \frac{d\kt^2}{\kt^2} \frac{\cf\as(\kt^2)}{\pi} \left[\log\frac{\mzz^2}{\kt^2}-\frac{3}{2}\right]\;.
\end{eqnarray}
The above formula has been obtained with the following assumptions
\begin{itemize}
\item The scale $Q$ is a scale of the order of the upper limit for $\kt$.
Its precise value affects the result only by terms of order $\as$, i.e.
next-to-next-to leading logarithmic (NNLL) terms.
\item We assume an implicit theta function associated to the pdf's,
that sets them to zero when the parton momentum fraction is greater than 1.
\item
We replace
\begin{equation}
\frac{d\kt^2}{\kt^2} = \frac{2y dy}{1-y^2} \to \frac{dy}{1\pm y}
\end{equation}
in the terms with no singularities in the $x$ integration,
and $y$ singularities in the
$y=\pm 1$ regions, the error for the replacement being of NNLL order.
\item
The upper limit of the $x$ integration can be set to 1 in the integrals that
do not have an $x\to 1$ singularity. In the terms singular for $x\to 1$,
we replace
\begin{equation}
 \sqrt{1-\frac{4x\kt^2}{(1-x)^2\mzz^2}}\to \sqrt{1-\frac{4\kt^2}{(1-x)^2\mzz^2}}\,,
\end{equation}
and thus set the upper limit of integration in $x$ to $1-2\kt/\mzz$, since this change makes
only subleading differences.
\item
We have used the leading order Altarelli-Parisi equations. Subleading
corrections to the evolution yield NNLL correction to the exponent.
\end{itemize}
We obtain
\begin{equation}\label{eq:sudll}
\Delta_q(\pt) \simeq \frac{f^{H_1}_q(x_{b1},\pt)}{f^{H_1}_q(x_{b1},Q)}
 \frac{f^{H_2}_{\bar{q}}(x_{b2},\pt)}{f^{H_2}_{\bar q}(x_{b2},Q)}
\exp\left\{
 -\int_{\pt^2}^{Q^2}
 \frac{d\kt^2}{\kt^2}  
\frac{\cf\as(\kt^2)}{\pi}\left[\log\frac{\mzz^2}{\kt^2}-\frac{3}{2}\right]
\right\}\;.
\end{equation}
Equation (\ref{eq:sudll}) corresponds to the NLL expression
of the Sudakov form factor in the DDT formulation \cite{Dokshitzer:1978hw}.
In fact, $\Delta_q$ multiplies the
Born term, that includes parton density functions evaluated at a scale of
the order of $Q$. The double ratio of parton density functions in
eq.~(\ref{eq:sudll}) thus replaces
the parton density functions included in the Born term with those evaluated
at the scale $\pt$, as required in the DDT formulation.
The exponent in eq.~(\ref{eq:sudll}) corresponds to the DDT exponent.
It is easy to check that, by replacing
\begin{equation}\label{eq:asimprov}
\as(\kt^2) \to \as(\kt^2) + \frac{1}{4\pi}
\left(\frac{67}{3}-\pi^2-\frac{10}{9}\nf\right)\,\as^2(\kt^2)
\end{equation}
in eq.~(\ref{eq:sudll})
we automatically achieve NLL accuracy in our Sudakov form factor\footnote{
In the language commonly used in Sudakov resummations, this amounts to the
inclusion of the $A_2$ term, that arises from the most singular contribution
of the next-to-leading order $P_{qq}$ splitting function.}
\cite{Ellis:1997ii, Frixione:1998dw}.
In fact, we may as well perform the replacement eq.~(\ref{eq:asimprov})
in our initial expression for $R$ in eq.~(\ref{eq:Deltaq}).
The remaining effects of such replacements in the derivation
carried above are in fact at the NNLL level.

The replacement eq.~(\ref{eq:asimprov}) was also advocated
in ref.~\cite{Catani:1990rr}, in the equivalent form of an
effective $\Lambda$ to be use in shower Monte Carlo programs,
$\Lambda_{\rm MC}=1.569\LambdaMSB $, in the framework of the
resummation of large $\log(1-x)$ effects in the threshold region.

\section{Hardest event generation} \label{sec:hardgen}
A straightforward implementation of the 
hardest event generation based upon the hardest event cross section
eq.~(\ref{eq:hardestsigma}) and standard Monte Carlo
techniques would not be very practical. In fact, the prefactor
$\bar{B}_q$ in eq.~(\ref{eq:hardestsigma}), as well as the Sudakov
exponent, already require a three dimensional integration
over the radiation variables. This may not be a severe problem
in the case of $Z$ pair production, since the production formulae
are relatively simple, but may become prohibitive for more
complex processes. In the following, we will illustrate a method
for generating hardest events with high efficiency. This method
is quite non-trivial, and it demonstrates the
applicability of the approach of ref.~\cite{Nason:2004rx} in the
most problematic case when initial state radiation is present.

For ease of presentation, we first illustrate how the
generation of hard events according to eq.~(\ref{eq:hardestsigma})
would proceed with standard Monte Carlo techniques.
The sequence is as follows:
\begin{enumerate}
\item\label{it:tot}
The total cross section $\sigtot$ is computed as
\begin{equation}
\sigtot=\sum_q \int d\Phi_v\, \bar{B}_q(v,\mu_v)\;.
\end{equation}
\item\label{it:genn}
Random values of $v$ and of the flavour label $q$
are generated with probability distribution $\bar{B}_q(v,\mu_v)$,
using standard hit-and-miss techniques,
schematically described in Appendix~\ref{app:hitandmiss}
(see also ref.~\cite{Dobbs:2004qw}).
\item
The radiation variables are generated as follows:
a real random number $0<n<1$ is generated uniformly, and
the equation
\begin{equation}
n=\Delta_q(\pt)
\end{equation}
is solved for $\pt$. If there is no solution (i.e., if the resulting
$\pt$ is below the infrared threshold) no radiation
is generated, and the event is produced as is. Otherwise,
radiation variables $r$ are generated with a distribution
proportional to
\begin{equation}
\delta(\kt-\pt) \, \frac{R_p(v,r,\mu_r)}{B_q(v,\mu_r)}\;,
\end{equation}
where $p$
(for ``process'') stands for $q\bar{q}$, $qg$, or $g\bar{q}$.
To be more specific one can use the $\delta$ function to eliminate
one variable, for example $x$, compute
\begin{equation}
D_p(v,y,\theta_2) \equiv
\int dx\, \delta(\kt-\pt) \, \frac{R_p(v,r,\mu_r)}{B_q(v,\mu_r)}
=\left.\frac{R_p(v,r,\mu_r)}{\frac{\partial \kt}{\partial x} B_q(v,\mu_r)}
\right\vert_{x=\bar{x}}\;,
\end{equation}
where $\bar{x}$ is such that $\kt(\bar{x},y,v)=\pt$,
and then generate $y$, $\theta_2$, and $p$ values distributed
with a probability proportional to $D_p(v,y,\theta_2)$ with
hit-and-miss techniques.
\end{enumerate}
The events generated according to the above prescription have
uniform weights, given by the total cross section computed
at step~\ref{it:tot} divided by the total number of generated
events.

The procedure outlined above is computationally intensive, both
in the generation of the Born variables, which requires one or more
three-dimensional integrations per generated point, and in the
generation of the radiation variables, where the computation of
the Sudakov exponent also requires a three-dimensional integration.
We shall now illustrate our method. We shall discuss separately the
generation of Born variables and the generation of radiation variables.

\subsection{Generation of the Born variables}\label{sec:borngen}
We first replace the
radiation variable $x$ by a rescaled variable $\hat x$
ranging between 0 and 1:
\be
\label{xhat}
\hat x=\frac{x-x_{\rm min}}{1-x_{\rm min}};\qquad
x = x_{\rm min}+\hat x(1-x_{\rm min}),
\ee
where $x_{\rm min}=x_{\rm min}(y,v)$ is given by eq.~(\ref{xmin}). 
In the two collinear regions $y=\pm 1$ we have, respectively,
$x_{\rm min}(1,v)=x_{b1}$ and $x_{\rm min}(-1,v)=x_{b2}$.
Therefore, we also define
\begin{equation}
x^+=x_{b1}+\hat x(1-x_{b1})\,,\quad
x^-=x_{b2}+\hat x(1-x_{b2})\,.
\end{equation}
The new radiation variables
${\hat r} \equiv \{{\hat x}, y,\theta_2\}$ have now fixed integration ranges
\begin{equation}
0<{\hat x}<1\,,\quad-1<y<1\,,\quad 0<\theta_2<\pi\,.
\end{equation}
We rewrite ${\bar B}_q(v,\mu)$ as
\begin{eqnarray}
{\bar B}_q(v,\mu)&=&B_q(v,\mu)+V_q(v,\mu)
\nonumber \\
&&+\int dy\, d\theta_2\, d\hat{x}\, (1-x_{\rm min})
 \,\left[ R_{q\bar{q}}(v,r,\mu)+R_{qg}(v,r,\mu)
+R_{g\bar{q}}(v,r,\mu)\right]
\nonumber \\
&&+\int d\hat{x}\,(1-x_{b1})
\left[ L^+_{q\bar{q}}(v,x^+,\mu)
+L_{g\bar{q}}(v,x^+,\mu)\right]
\nonumber \\
&&+\int d\hat{x}\, (1-x_{b2})
 \left[ L^-_{q\bar{q}}(v,x^-,\mu)
+L_{qg}(v,x^-,\mu)\right]
\nonumber \\
&=&\int dy\,d\theta_2\,d{\hat x}\, {\tilde B}_q(v,\hat{r},\mu)\,,
\end{eqnarray}
where 
\begin{eqnarray}
{\tilde B}_q(v,\hat{r},\mu)&=&
\frac{1}{2\pi}\left[B_q(v,\mu)+V_q(v,\mu)\right]
\nonumber \\ &&
+ (1-x_{\rm min})
 \,\left[ R_{q\bar{q}}(v,r,\mu)+R_{qg}(v,r,\mu)
+R_{g\bar{q}}(v,r,\mu)\right]
\nonumber \\
&&+\frac{1}{2\pi} (1-x_{b1})
\left[ L^+_{q\bar{q}}(v,x^+,\mu)
+L_{g\bar{q}}(v,x^+,\mu)\right]
\nonumber \\
&&+\frac{1}{2\pi} (1-x_{b2}) \left[ L^-_{q\bar{q}}(v,x^-,\mu)
+L_{qg}(v,x^-,\mu)\right],
\end{eqnarray}
and we define
\begin{equation}
{\tilde B}(v,\hat{r},\mu)=\sum_q {\tilde B}_q(v,\hat{r},\mu)\;,
\end{equation}
so that
\begin{equation}
\sigtot = \int d\Phi_{\hat r}\,d\Phi_v\, {\tilde B}(v,\hat{r},\mu_v)\,,
\label{sigmatot}
\end{equation}
where $d\Phi_{\hat r}= dy \,d\theta_2 \,d {\hat x}$.

We now observe that the integrand in eq.~(\ref{sigmatot}) is, in general,
positive. It is in fact the sum of a positive term of order 0
in $\as$ and terms of order $\as$. Negative terms are not totally
forbidden, but they clearly disappear in the perturbative (i.e.
$\as \to 0$) limit. It will be thus necessary, when performing
the numerical integration, to check that the contribution of negative terms
is negligible.

It is possible to store certain intermediate results of the numerical
integration procedure, that can be subsequently used to generate
efficiently the variables $v,{\hat r}$ with a distribution
proportional to ${\tilde B}(v,\hat{r},\mu_v)$.  We give in
appendix~\ref{app:unweighting} an elementary description of such a
procedure.  In practice, computer programs are available that
implement and optimize this procedure.  We have used the BASES/SPRING
\cite{Kawabata:1995th} package.  The adaptive Monte Carlo integration
routine BASES performs the integration and stores the necessary
intermediate results.  The routine SPRING uses this information to
generate unweighted events according to the distribution ${\tilde
B}(v,{\hat r},\mu_v)$. In this way, both $v$ and $\hat{r}$ variables
are generated, and then the flavour $q$ of the event is generated with
a probability proportional to ${\tilde B}_q(v,\hat{r},\mu_v)$.  The
values of the radiation variables ${\hat r}$ are then ignored, since
we are only interested in the distribution of the Born variables $v$.
This precisely amounts to integrating away the ${\hat r}$ variables,
so that we are left with a uniform generation of $v$ and $q$ values
according to the ${\bar B}_q(v,\mu)$ distribution.

\subsection{Generation of the radiation variables}\label{sec:radgen}
Having generated the Born variables $v$ and the flavour $q$,
the next step is the generation of the radiation variables
according to the probability distribution
\begin{equation}
\Delta_q(v,\kt(v,r))\, W_q(v,r)\, d\Phi_r\;,
\end{equation}
where
\ba \label{eq:Wdef}
&&W_q(v,r)=
 \frac{{\hat R}_{q\bar{q}}(v,r,\mu_r)+{\hat R}_{qg}(v,r,\mu_r)
+{\hat R}_{g\bar{q}}(v,r,\mu_r)}{B_q(v,\mu_r)}
\\
&&
\Delta_q(\pt,v)=\exp\left[-\int W_q(v,r^\prime)\,\theta(\kt(v,r^\prime)-\pt)\,
 d\Phi_{r^\prime} \right]\;.
\ea
We do this by means of the veto technique, described in some detail in
appendix~\ref{app:veto}.
We find a function $U_q(v,r)$ such that
\begin{equation}
U_q(v,r)\ge W_q(v,r)\;,
\end{equation}
and we generate radiation variables $r$ according to the
distribution
\begin{equation}
\label{eq:distrUq}
\Delta_q^{(U)}(v,\kt(v,r)) \,U_q(v,r)\, d\Phi_r
\end{equation}
where
\begin{equation}
\Delta_q^{(U)}(v,\pt)=\exp\left[-\int U_q(v,r^\prime)\,
 \theta(\kt(v,r^\prime)-\pt)\, d\Phi_{r^\prime}\right]\,.
\end{equation}
The generation of the event is then performed by the following steps:
\begin{enumerate}
\item
Set $p_{\rm max}=\infty$.
\item \label{it:ptloop}
Generate a random number $n$ between 0 and 1, and solve the equation
\begin{equation}\label{eq:genU}
n=\frac{\Delta_q^{(U)}(v,\pt)}{\Delta_q^{(U)}(v,p_{\rm max})}
\end{equation}
for $\pt$.
\item \label{it:rgen}
Generate the variables $r$ according to the distribution
\begin{equation}
U_q(v,r)\,\delta(\kt(v,r)-\pt)\;.
\end{equation}
\item
Accept the generated value with a probability
$W_q(v,r)/U_q(v,r)$. If the event is rejected, set $p_{\rm max}=\pt$,
and go to step \ref{it:ptloop}.
\end{enumerate}
It is obvious that an efficient generation is achieved if
the chosen functional form for $U_q$ is such that the solution of
eq.~(\ref{eq:genU}) and the generation of $r$ variables at step \ref{it:rgen}
are simple, and that the ratio $W_q(v,r)/U_q(v,r)$ is
not too far below 1 in most of the integration range. 
We find that the choice 
\begin{equation}\label{eq:Maggiorante}
U_q(v,r)=N_q \, \frac{\as(\kt^2(v,r))}{(1-x)\,(1-y^2)}\,,
\end{equation}
with a suitable choice of the constant  $N_q$, fulfills both requirements.
The generation of events according to the distribution in
eq.~(\ref{eq:distrUq}), with $U_q(v,r)$ given by eq.~(\ref{eq:Maggiorante}),
is not entirely trivial; we describe it in Appendix~\ref{app:Uq}.

The choice in eq.~(\ref{eq:Maggiorante}) is suggested by the structure of the
function $W_q$ in eq.~(\ref{eq:Wdef}) near the collinear limit,
as we now show.
Let us consider first the $q\bar{q}$ contribution.
In the positive collinear direction ($y \to 1$) we have
(see eq.~(\ref{eq:singlimit}))
\begin{equation}
\frac{{\hat R}_{q\bar{q}}(v,r,\mu_r)}{B_q(v,\mu_r)}
\rightarrow \frac{1}{\pi} \frac{1}{1-y} \frac{C_F \as}{2\pi} \frac{1+x^2}{1-x}
\;\frac{f^{H_1}_q(x_{b1}/x,\mu_r)}{x\,f^{H_1}_q(x_{b1},\mu_r)}\;.
\label{eq:collim}
\end{equation}
A similar relation holds for $y\to -1$.
It is now reasonable to assume that the parton density ratio in
eq.~(\ref{eq:collim}) is
bounded by a number of order one, since parton densities, in general,
are never fast growing functions of $x$. A bounding function of
the form of eq.~(\ref{eq:Maggiorante}) therefore arises in a natural
way.
We now consider the $g{\bar{q}}$ contribution. In the $y\to 1$ limit
we have
\begin{equation}
\frac{{\hat R}_{g\bar{q}}(v,r,\mu_r)}{B_q(v,\mu_r)}
\rightarrow \frac{1}{\pi} \frac{1}{1-y} \frac{T_F \as}{2\pi}
\left[(1-x)^2+x^2\right]
\;\frac{f^{H_1}_g(x_{b1}/x,\mu_r)}{x\,f^{H_1}_q(x_{b1},\mu_r)}\;.
\end{equation}
In this case, the ratio of parton densities 
is between different parton species, and must be discussed with care.
First of all, we notice that 
\be
f^{H_1}_g(x_{b1}/x,\mu_r)\lesssim f^{H_1}_g(x_{b1},\mu_r)\,,
\ee
so that we only need a bound for
\be
\frac{f^{H_1}_g(x_{b1},\mu_r)}{f^{H_1}_q(x_{b1},\mu_r)}\,.
\ee
At small values of $x_{b1}$, this ratio is always bounded, because
the gluon and quark densities have similar behaviour in the
small-$x$ limit. For large $x_{b1}$, if
$q$ is a valence quark, the ratio is also bounded, since the
gluon is generated by valence quarks through evolution.
In case $q$ is a sea quark, the corresponding density may be
softer than the gluon density. In the worst case, however,
the sea quark is generated by the gluon through evolution.
Assuming that parton densities
behave as a power of $1-x$ at large $x$, 
\begin{equation}
f_g(x,\mu) \sim (1-x)^\delta
\end{equation}
the Altarelli-Parisi equation in the large $x$ limit yields
\begin{equation}
\mu^2 \frac{d f_q(x,\mu)}{d\mu^2} \sim \frac{T_F \as}{2\pi}
\int_x^1\frac{dz}{z}\, (1-z)^\delta
\sim \frac{T_F \as}{2\pi}  (1-x)^{\delta+1}\;,
\end{equation}
and therefore
\be
\frac{f^{H_1}_g(x_{b1},\mu_r)}{f^{H_1}_q(x_{b1},\mu_r)}
\lesssim \frac{1}{1-x_{b1}}\,.
\ee
Since
\be
\frac{1}{1-x_{b1}}<\frac{1}{1-x}\;,
\ee
we conclude that the choice eq.~(\ref{eq:Maggiorante})
is adequate also in this case. 

The normalization factor $N_q$ can be obtained numerically, using
SPRING to generate $v,r$ and $q$ values, and then computing the
maximum of the ratio $U_q(v,r)/W_q(v,r)$.
\subsection{Colour assignement}
The NLO calculation of a generic production process may specify also
the detailed colour structure of the produced particles.  SMC
generators use the colour information for the subsequent shower, and
also at the hadronization stage, for the formation of colour singlet
hadrons. In general, only the colour flow structure in the limit of a
large $\Nc$ (where $\Nc$ is the number of colours) is needed, and, in
fact, in the Les Houches Interface one can only specify the colour
connections of the intervening partons.  In the case of $ZZ$
production, the large $\Nc$ colour assignment is the same in all real
emission graphs, corresponding to the quark and antiquark colour line
merging into the gluon double line. This is thus the colour structure
that must be passed to the LHI.

In the general case, several colour configurations are possible, and
one should specify which one to choose after the radiation has been
generated.  If the contribution to the real emission cross section is
available for each (large $N_c$) colour component, one simply chooses
the colour component with a probability proportional to it.

\section{Results and comparisons}
\label{sec:results}
We now illustrate some numerical results obtained by
the hardest event generation prescription presented
in the previous Sections. We will refer to this method as
as the \POWHEG{} (for Positive weight Hardest event Generator).
Our results are obtained with the following default choices:
\begin{itemize}
\item
We use the CTEQ6M~\cite{Pumplin:2005rh} parton density set.
\item
We fix the factorization and renormalization scales in the
computation of $\bar{B}_q$ to the invariant mass of the
$Z$ boson pair, $\mu_v=\mzz$.
\item
We fix the factorization and renormalization scales for radiation
to the $\kt$ of the radiated parton, eq.~(\ref{eq:ktdef}).
\item
We use the value of $\LambdaMSB$ appropriate to the PDF set we
have chosen. This value is corrected according to the
prescription given in Section~\ref{sec:hard} when generating
radiation variables.
\item
We consider two experimental configurations of interest:
$p\bar{p}$ collisions at $\sqrt{S}=1920\;$GeV, corresponding
to the Tevatron, and $pp$ collisions at $\sqrt{S}=14\;$TeV,
corresponding to the LHC.
\item
We use a fixed number of flavours $\nf=5$. In principle this choice
is not completely consistent. One should instead reduce $\nf$
when generating radiation below the bottom and charm scales. This has
however a minor impact on the final results, and we have chosen not to
take it into account at this stage.
\item
The $Z$ bosons are treated as stable particles,
with $\mz=91.2\;$GeV. We have forced, whenever possible,
the same assumption on standard Monte Carlo predictions,
when comparing them to our results.
\item
We have used the following values for the electroweak parameters:
$\sin^2\theta_{\sss W}=0.23113$
and $\alpha_{\rm em}=\alpha_{\rm em}(\mz^2)\simeq 1/128$.
\end{itemize}

\begin{figure}[ht]
\begin{center}
\epsfig{file=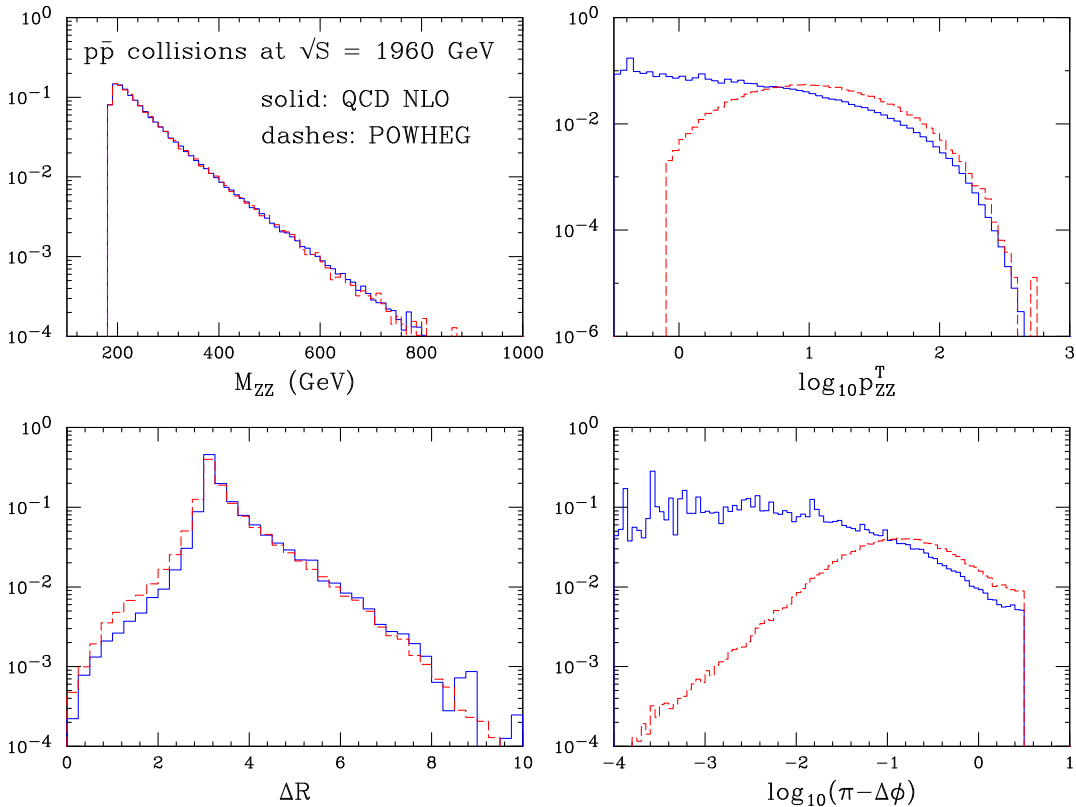,width=0.95\textwidth}
\end{center}
\caption{\label{fig:fo-mcpnlo-tev}
Comparison of four distributions computed according to the fixed
order calculation and the \POWHEG{}.
On the vertical axes we report the cross sections in picobarns per bin.
Energies are expressed in GeV.}
\end{figure}

We begin by comparing the \POWHEG{} results with the fixed order
QCD calculation \cite{Mele:1990bq}, performed with the same scale choice
adopted for $\bar{B}_q$, $\mu=\mzz$. We have considered distributions
in the following variables: the transverse momentum and rapidity of
one $Z$ boson, the rapidity and pseudorapidity differences $\Delta y$
and $\Delta \eta$ between the two $Z$'s, the pair invariant mass
$\mzz$, the transverse momentum and rapidity of the $ZZ$ pair, the
azimuthal distance $\Delta\phi$ between the two $Z$'s, and $\Delta
R=\sqrt{\Delta \phi ^2 + \Delta \eta^2}$. 
Among the distributions we have considered, the $\Delta \phi$ and $\ptzz$
distributions are strongly affected by light parton emission,
since they are trivial in the case of no emission. The $\Delta R$
distribution is an intermediate case, since it depends upon $\Delta \phi$.
All other variables are non-trivial already at the Born level.
We find that the NLO calculation and \POWHEG{} give equivalent results
for this last group of observables, while in the case
of  $\Delta \phi$ and $\ptzz$, and, to a lesser extent, of  $\Delta R$,
we find important differences.
This is illustrated in fig.~\ref{fig:fo-mcpnlo-tev},
where the  $\Delta \phi$, $\ptzz$ and $\Delta R$
distributions are shown, together with the invariant mass distribution,
which is taken as a representative example of all the remaining observables.
We first notice that in the case of the invariant mass distributions
the two calculations give identical results.
On the contrary, the $\Delta \phi$ and $\ptzz$
distributions are strongly affected by light parton emission, and indeed
they display a sizable difference in the two calculations.  In the fixed order
calculation the divergences in the limit of soft/collinear light
parton emissions are canceled when virtual corrections, that have
strictly $\ptzz=0$ and $\Delta \phi=\pi$, are included.  However, the
effect of virtual corrections is not seen in the $\Delta \phi$ and
$\ptzz$ plots, so that the fixed order calculation appears to yield
logarithmically divergent integrals for $\ptzz \to 0$ and $\Delta \phi
\to \pi$.  On the other hand, the \POWHEG{} result is well behaved also
in this region, since the no-radiation region is effectively
suppressed by the form factor $\Delta_q(\kt)$ (see
eq.~(\ref{eq:hardestsigma})).

\begin{figure}[ht]
\begin{center}
\epsfig{file=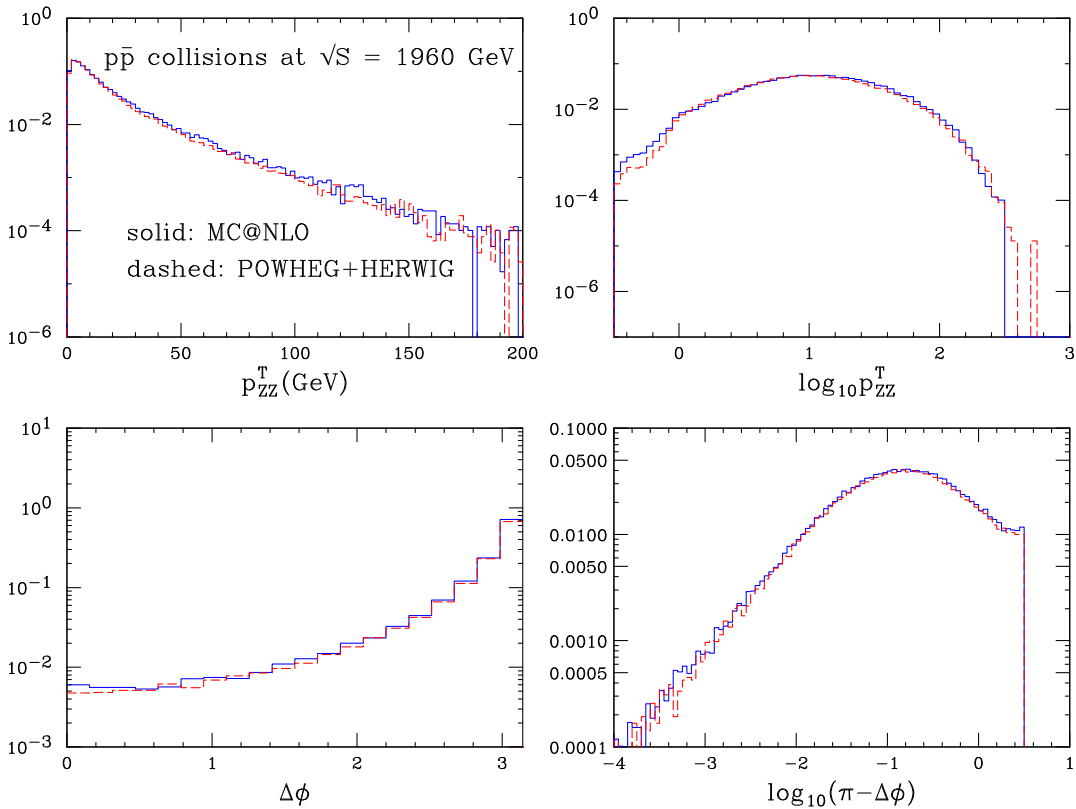,width=0.95\textwidth}
\end{center}
\caption{\label{fig:mcatnlo-pheg-tev-1}
Comparison of four distributions computed according to \MCatNLO{}
and the \POWHEG{}.
On the vertical axes we report the cross sections in picobarns per bin.
Energies are expressed in GeV.}
\end{figure}

\begin{figure}[ht]
\begin{center}
\epsfig{file=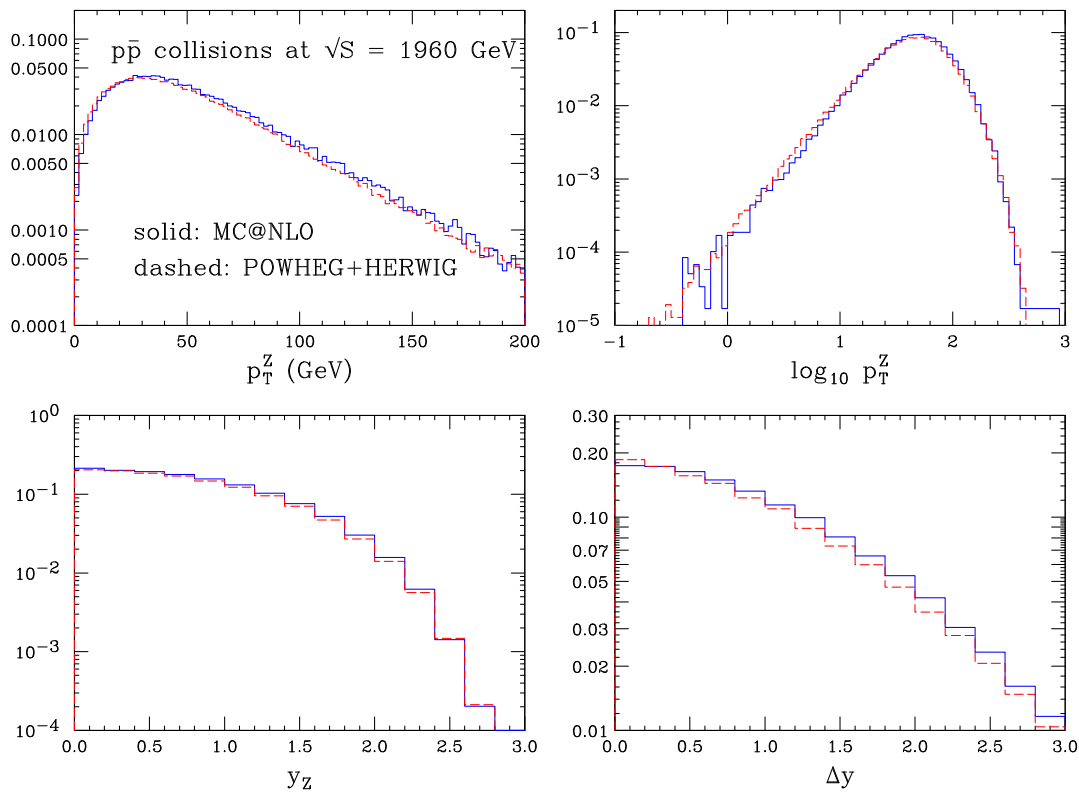,width=0.95\textwidth}
\end{center}
\caption{\label{fig:mcatnlo-pheg-tev-2}
Comparison of four distributions computed according to \MCatNLO{}
and the \POWHEG{}.
On the vertical axes we report the cross sections in picobarns per bin.
Energies are expressed in GeV.}
\end{figure}

\begin{figure}[ht]
\begin{center}
\epsfig{file=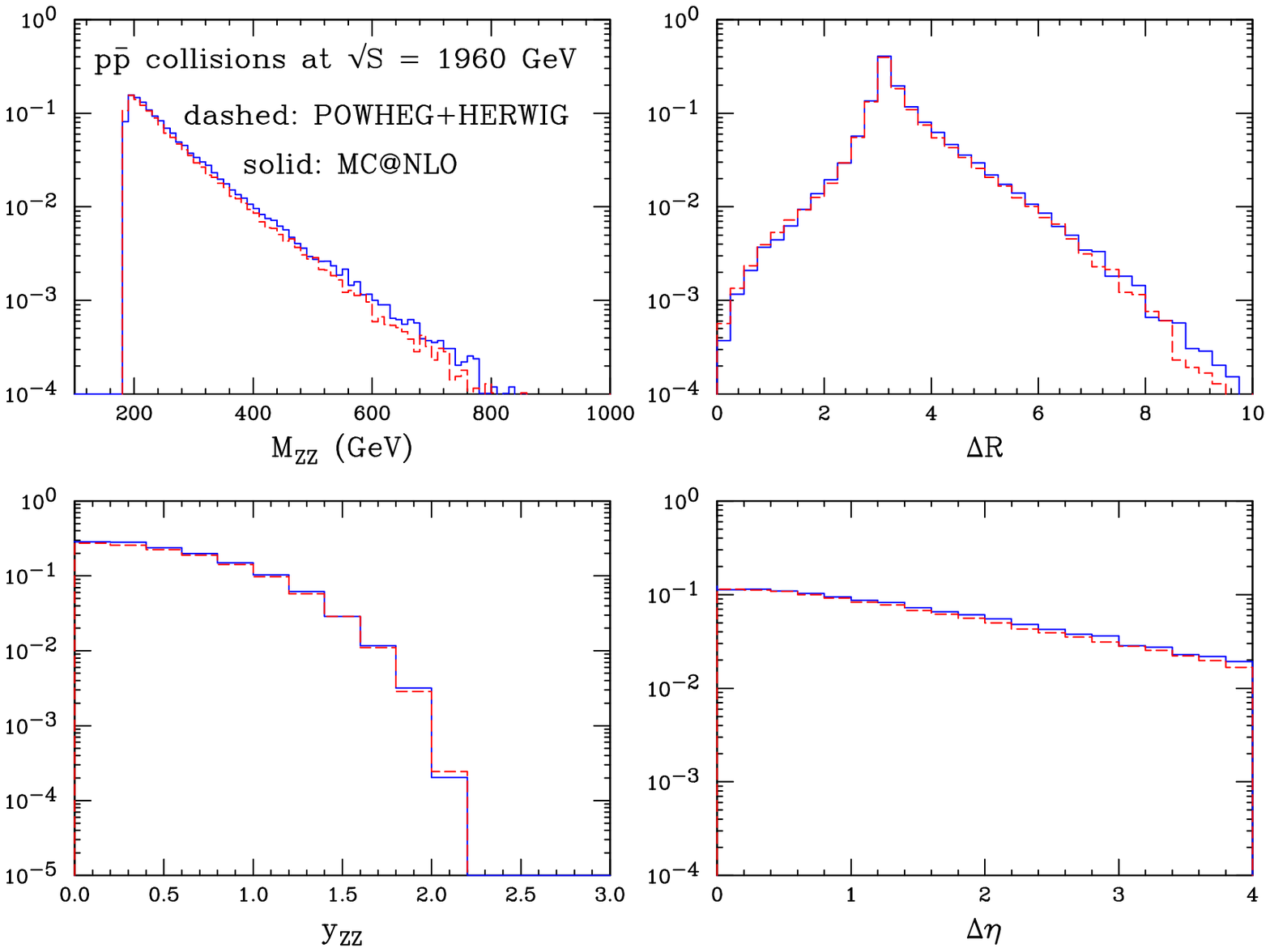,width=0.95\textwidth}
\end{center}
\caption{\label{fig:mcatnlo-pheg-tev-3}
Comparison of four distributions computed according to \MCatNLO{}
and the \POWHEG{}.
On the vertical axes we report the cross sections in picobarns per bin.
Energies are expressed in GeV.}
\end{figure}

\begin{figure}[ht]
\begin{center}
\epsfig{file=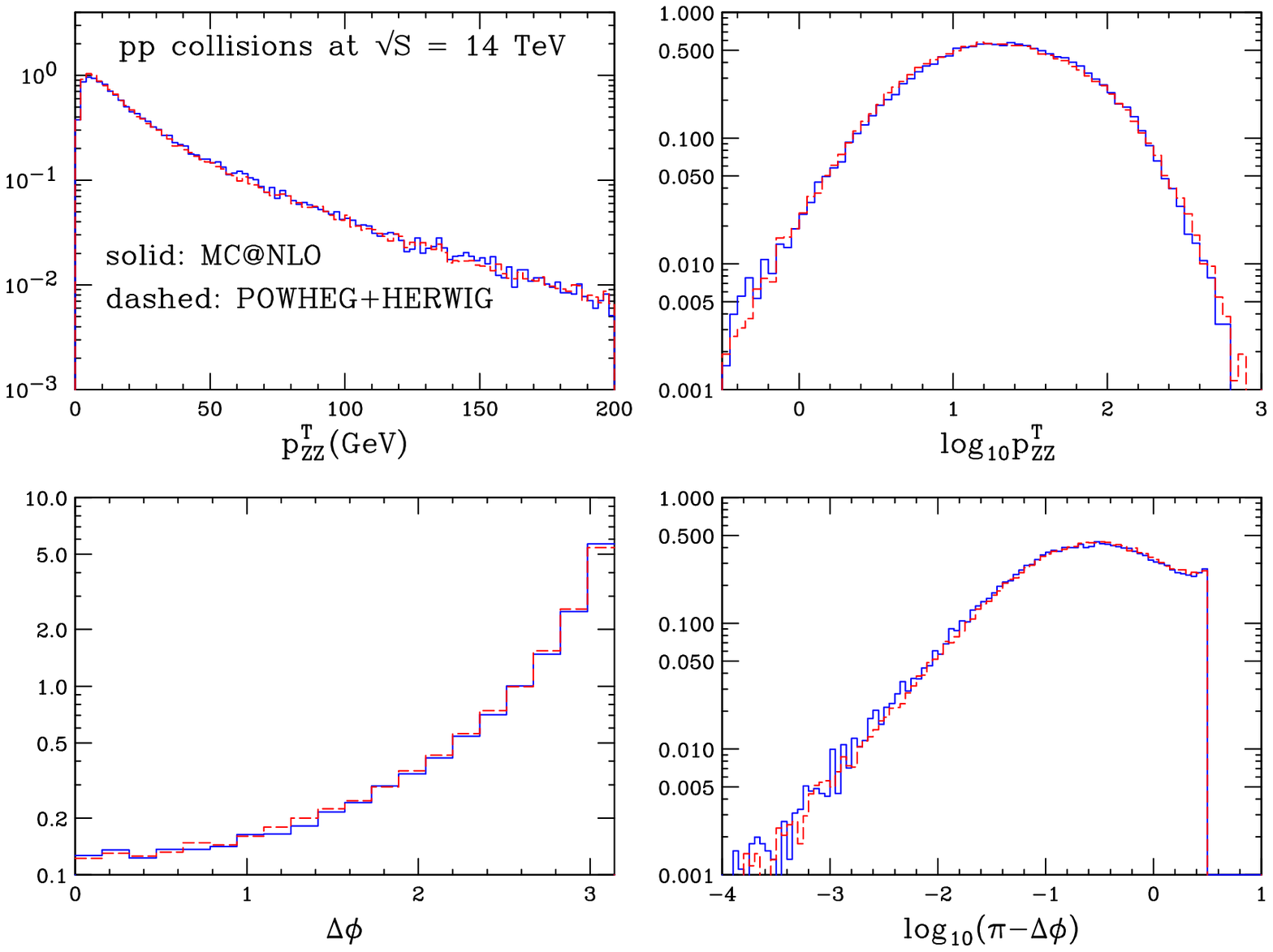,width=0.95\textwidth}
\end{center}
\caption{\label{fig:mcatnlo-pheg-lhc-1}
Same as fig.~\ref{fig:mcatnlo-pheg-tev-1} for the LHC.}
\end{figure}

\begin{figure}[ht]
\begin{center}
\epsfig{file=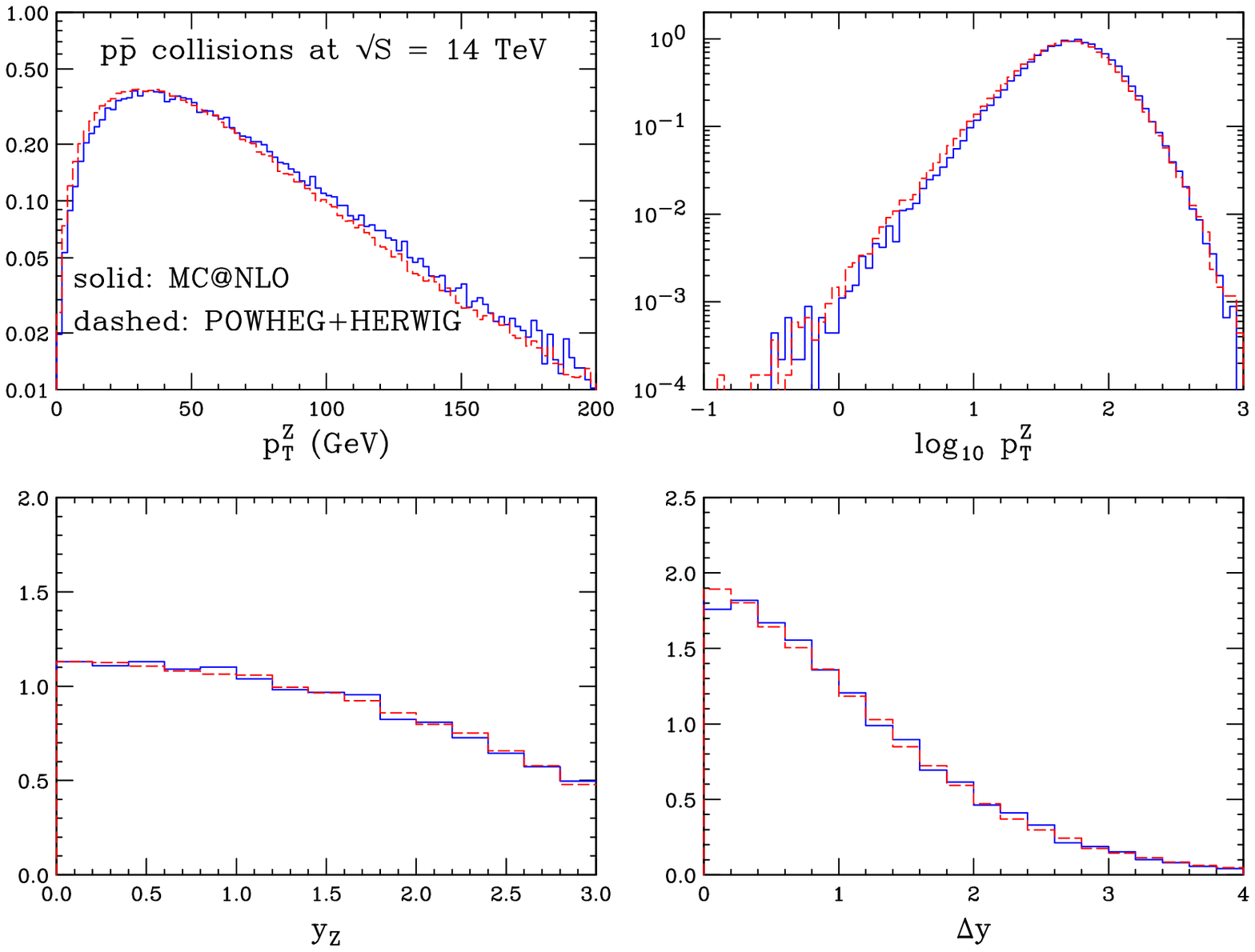,width=0.95\textwidth}
\end{center}
\caption{\label{fig:mcatnlo-pheg-lhc-2}
Same as fig.~\ref{fig:mcatnlo-pheg-tev-2} for the LHC.}
\end{figure}

\begin{figure}[ht]
\begin{center}
\epsfig{file=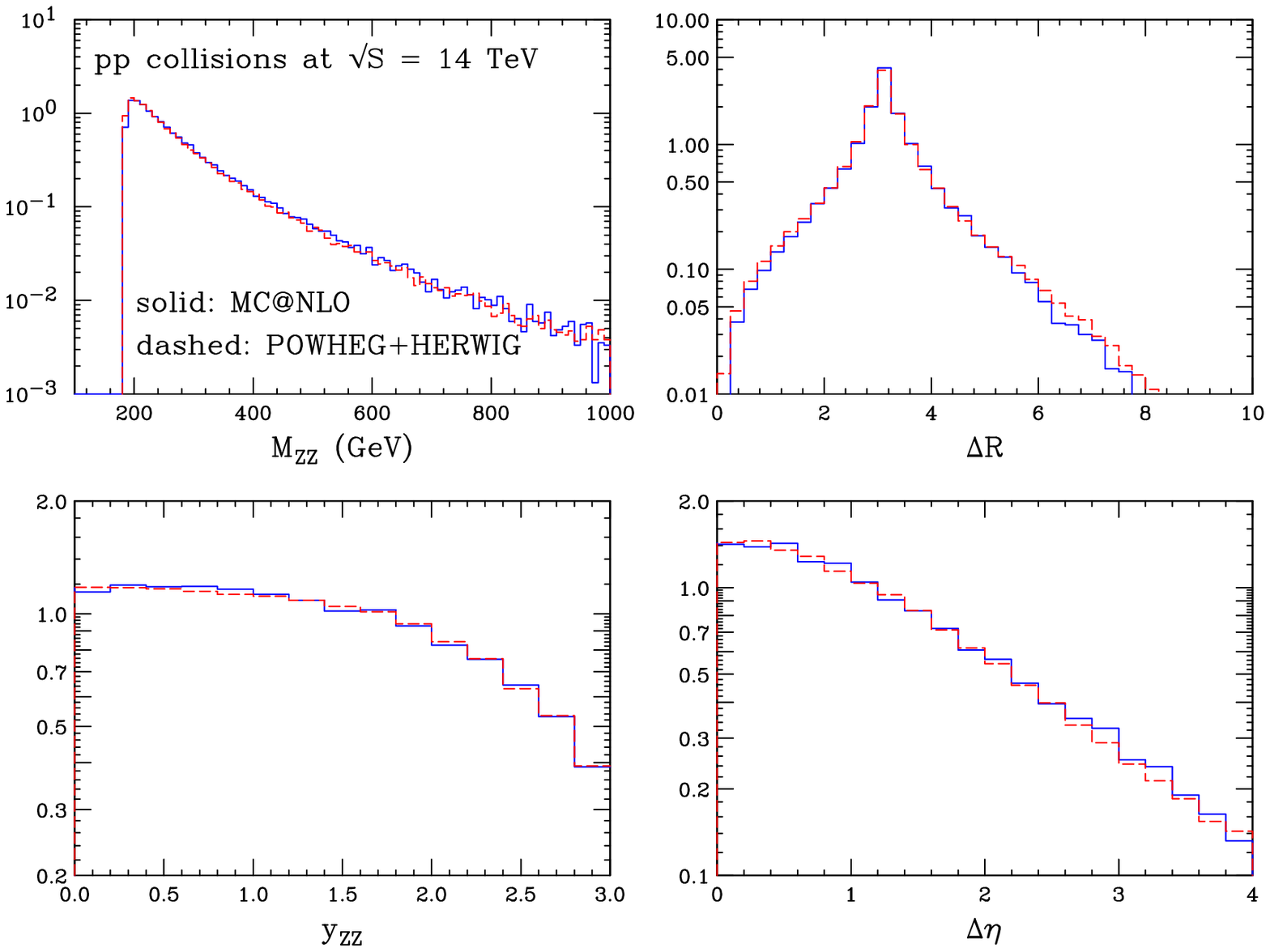,width=0.95\textwidth}
\end{center}
\caption{\label{fig:mcatnlo-pheg-lhc-3}
Same as fig.~\ref{fig:mcatnlo-pheg-tev-3} for the LHC.}
\end{figure}

Our next task is to compare our full result with the only existing
implementation of Monte Carlo generation of vector boson
pairs with NLO accuracy, namely the \MCatNLO{} program.
The relevant plots are collected in figs.~\ref{fig:mcatnlo-pheg-tev-1},
\ref{fig:mcatnlo-pheg-tev-2}, \ref{fig:mcatnlo-pheg-tev-3} for the Tevatron,
and in figs. \ref{fig:mcatnlo-pheg-lhc-1},
\ref{fig:mcatnlo-pheg-lhc-2}, \ref{fig:mcatnlo-pheg-lhc-3} for the LHC.
The \MCatNLO{} results are compared with \POWHEG{} interfaced
with HERWIG. According to the Les Houches interface prescription,
the showering in the Monte Carlo is vetoed by assigning the
$\kt$ of the event generated by \POWHEG{} to the variable
{\tt SCALUP} in the Les Houches common block. We find
that the inclusion of the HERWIG shower determines only tiny
changes with respect to the pure \POWHEG{} output.
It is easy to comment upon the outcome of this comparison:
the two algorithms yield identical results.
Minor differences are only seen in the $\ptzz$
distribution; these can be easily attributed to the
presence of a $\kt$ hard cutoff, set at 0.8~GeV in the \POWHEG{}.

It should be noted that the \MCatNLO{} results
have been computed using the scale choice
suggested by the authors, namely
\be
\mu^2=\frac{1}{2}\left(\sqrt{p_{T1}^2+\mz^2}+\sqrt{p_{T2}^2+\mz^2}\right),
\ee
while the \POWHEG{} results are obtained with the scale choices described
at the beginning of this section. This difference has a negligible
impact on the total cross section, but may affect some of the
distributions.
The agreement between the two methods in the region of large
transverse momentum radiation is expected, since both methods are in
good agreement with the fixed-order calculation in this region.  The
region of soft radiation in the \MCatNLO{} implementation is
controlled by the HERWIG shower. We therefore conclude that our
treatment of the soft region is consistent with the HERWIG
implementation.  We point out that HERWIG fully implements the
procedure discussed at the end of Section~\ref{sec:hard}
for the treatment of the
strong coupling constant in the shower emission, as we also do.

\begin{figure}[ht]
\begin{center}
\epsfig{file=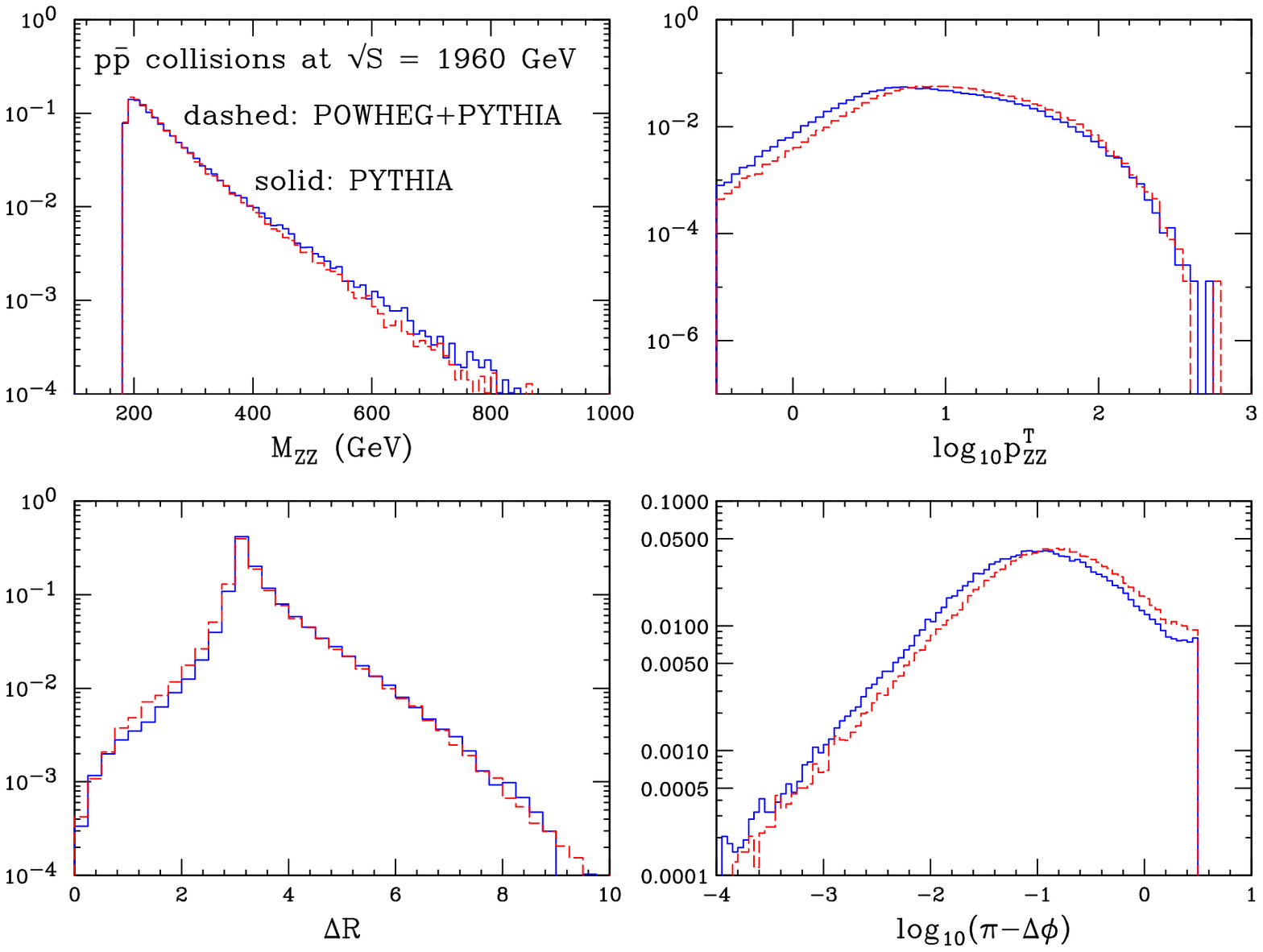,width=0.95\textwidth}
\end{center}
\caption{\label{fig:py-pheg-py-tev-1}
Comparison of four distributions computed according to PYTHIA{}
and the \POWHEG{} interfaced to PYTHIA.}
\end{figure}

\begin{figure}[ht]
\begin{center}
\epsfig{file=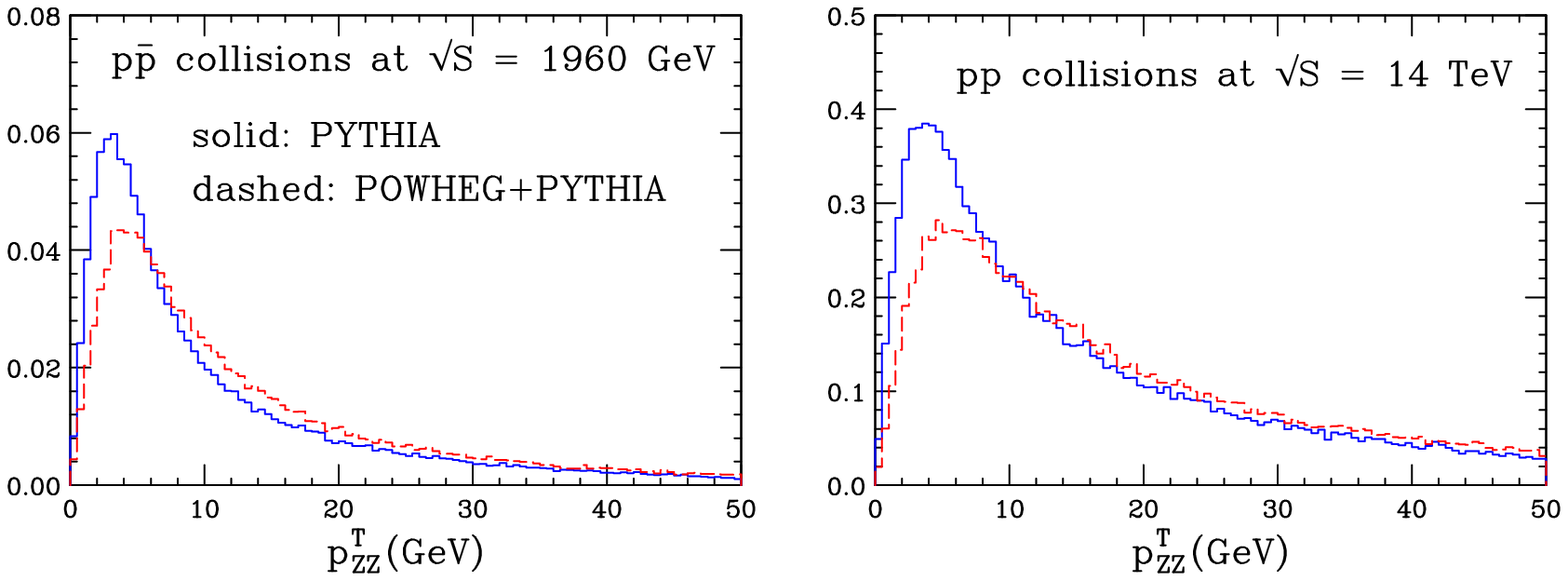,width=0.95\textwidth}
\end{center}
\caption{\label{fig:py-pheg-ptpair}
Comparison of the $\ptzz$ distribution
computed according to PYTHIA
and the \POWHEG{} interfaced to PYTHIA, for both the
Tevatron and the LHC.}
\end{figure}

One of the main features of the \POWHEG{} method is the possibility of
interfacing its output to any shower Monte Carlo that implements the
Les Houches interface for user-provided processes. This is possible
with the popular Monte Carlo PYTHIA~\cite{Sjostrand:2006za}
since its version 6.3, which is based upon
a $\pt$-ordered shower algorithm. We thus interfaced \POWHEG{} to PYTHIA
and compared the results to the pure PYTHIA output, normalized to
the \POWHEG{} cross section. 
By inspection of fig.~\ref{fig:py-pheg-py-tev-1}, relevant to the Tevatron
configuration,
we see that the \POWHEG{} result appears to have a harder
$\ptzz$ spectrum, especially in the low $\ptzz$ region.
Correspondingly, the $\Delta\phi$ distribution obtained
by \POWHEG{} is also shifted away from the region $\Delta\phi\sim\pi$
with respect to the PYTHIA result.
A detailed view of the small-$\ptzz$ region is shown
in fig.~\ref{fig:py-pheg-ptpair}
for both the Tevatron and the LHC.
We see that the
position of the peak in the $\ptzz$ distribution is rather different
in the two approaches. We have verified that also in this case,
the effect of the PYTHIA showering in the \POWHEG{}$+$PYTHIA result is
negligible in the distributions we are considering. We thus infer that
the same differences should appear when comparing PYTHIA with \MCatNLO{}.
This issue deserves further investigations, that are however beyond
the purposes of the present work.

\section{Conclusions}
\label{sec:conclusions}
We have presented an explicit implementation of the method presented
in ref.~\cite{Nason:2004rx} for the construction of a Monte Carlo
event generator with matrix elements accurate to next-to-leading
order, and positive weights. The process we have considered is $Z$ pair
production in hadron collisions, which is a process of intermediate
complexity that involves initial-state
radiation, and thus is a good testing ground for the applicability
of the method to hadron collision processes.

One of the main achievements of the present work is
the development of numerical techniques that are necessary
for the practical implementation of the method proposed
in ref.~\cite{Nason:2004rx}.
We have given here a full, detailed description of these techniques,
so that all our results can in principle be reproduced by
the interested reader.

The output of our generator is cast in the form of the Les Houches
Interface for user provided processes~\cite{Boos:2001cv},
and thus easily interfaced to both the HERWIG and the PYTHIA
shower Monte Carlo programs.

We have compared our result to the only existing
program that can compute NLO corrected Shower Monte Carlo
output in hadronic collisions\footnote{Other methods for NLO implementations
for shower Monte Carlo have been presented in the literature \cite{Nagy:2005aa},
that are limited to the case of $e^+e^-$ annihilation, and do not have positive
weights.},
for a large set of observables,
both for the Tevatron and the LHC. We have found an excellent
agreement.

Our method can also be used as a standalone, alternative implementation
of QCD corrections. As discussed in section~\ref{sec:results},
the interfacing to the SMC
has negligible effects on the \POWHEG{} distributions involving
$ZZ$ observables. It may thus becomes convenient to compute these
observable by using \POWHEG{} as a standalone program. This has several advantages
over the standard, fixed order QCD calculation, like positivity,
and next-to-leading logarithmic accuracy in the small $\kt$ region. 

The extension of the present work to $WZ$ and $WW$ pair production
is straightforward. Furthermore, the choice of kinematic variables
adopted here can
be easily extended to processes of Drell-Yan type, like
single $W$ and $Z$ production, and Higgs production.
We also believe that the method can be applied without major
modifications to the case of heavy flavour production, where
final state soft radiation is also present.
The treatment of collinear final state radiation with the
method of ref.~\cite{Nason:2004rx} presents no difficulties,
being in fact easier than the initial state radiation case.
The formulation of the application of the method to general
processes is under study.

In order to fully preserve double logarithmic accuracy in the
\POWHEG{} context, the Shower Monte Carlo to which the program
is interfaced should generate consistently the needed soft
radiation pattern. When using an angular ordered shower program,
one needs to include additional soft radiation, that was
named ``truncated showers'' in ref.~\cite{Nason:2004rx}.
Programs that generate soft radiation using dipole type formalisms
\cite{Gus88,Pet88}
may not require any further correction. At present, however,
no such program is available that handles initial state radiation.
These problems are left for future studies.

\vspace{12pt}
\noindent
{\bf Acknowledgments}:
We thank Stefano Frixione for helpful
discussions and suggestions, and Carlo Oleari for carefully reading
the manuscript.
We also thank Torbj\"orn Sj\"ostrand for clarifications on the use of PYTHIA.

\appendix
\section{The hit-and-miss technique}
\label{app:hitandmiss}
In order to generate a set of continuous variables $x$
and an index $j=1,\ldots,m$, with a probability distribution
proportional to $F_j(x)$,
one first computes
\begin{equation}
F_{\rm tot} = \sum_j \int dx\, F_j(x)\,,\quad\quad F_{\rm max} =
\max_x \sum_j F_j(x)\;.
\end{equation}
Then, one generates a random value $x={\bar x}$ and
a random number $0<n<1$ with flat distributions,
and finds the smallest value $k$ such that
\begin{equation}
\sum_{j=1}^k F_j({\bar x})> n\,F_{\rm max} \;.
\end{equation}
If such $k$ does not exist, the event is rejected; otherwise
the set $j=k,x={\bar x}$ is accepted.

\section{From Monte Carlo integration to uniform generation}
\label{app:unweighting}
We want to to generate a set of values for the variables $x$ in a domain
${\cal D}$ with a distribution proportional
to $F(x)$. We divide the integration range ${\cal D}$ into
a set of hypercubes ${\cal D}_i$, with $i=1,\ldots, m$.
For each hypercube we compute
\begin{equation}
F_{\rm tot}^{(i)}=\int_{{\cal D}_i} dx\, F(x)\,,\quad\quad
F_{\rm max}^{(i)}=\max_{x \in {\cal D}_i} F(x)\,,\quad\quad
F_{\rm tot}=\sum_{i=1}^m F_{\rm tot}^{(i)}\;.
\end{equation}
In order to generate the $x$ values, we first choose the hypercube
according to its probability: given a random number $0<n<1$
we find the minimum value $k$ such that
\begin{equation}
\sum_{j=1}^k F^{(j)}_{\rm tot}> n\,F_{\rm tot}\;.
\end{equation}
Next we generate $x$ in the ${\cal D}_k$ hypercube using the
hit-and-miss technique. It is clear that the efficiency of the
generation improves by reducing the size (and thus increasing
the number) of the hypercubes. Thus there will be a trade-off
between speed and storage requirement in the computer implementation
of this technique.

\section{The veto technique}
\label{app:veto}
Assume we want to generate values for a set of variables
$x$, according to a distribution
\begin{equation}\label{eq:distr}
P(x)=R(x)\,
\exp\left[-\int d^dx^\prime\, R(x^\prime)\, \theta(p(x^\prime)-p(x))\right]\,,
\end{equation}
where $p(x)\geq 0$.  We assume that $R(x)$ is non-negative, and that
the unconstrained integral is divergent.  We observe that the
probability distribution of $p(x)$ is an exact differential.  Indeed,
\begin{eqnarray}
\int d^d x\, P(x)\,\delta(p(x)-p)
&=&\int d^dx \,R(x)\,\delta(p(x)-p)\,
\exp\left[-\int d^dx^\prime\, R(x^\prime)\,
\theta(p(x^\prime)-p)\right]
\nonumber\\
&=&\frac{d}{dp} \Delta(p)\;,
\label{eq:exactdiff}
\end{eqnarray}
where we have defined
\begin{equation}
\label{eq:deltadef}
\Delta(p)=\exp\left[-\int d^dx^\prime\, R(x^\prime)\,
\theta(p(x^\prime)-p)\right]\;.
\end{equation}
The function $\Delta(p)$ ranges between 0 and 1, because
the integral in the exponent vanishes for $p\to+\infty$ and diverges
to $+\infty$ for $p=0$. Hence, eq.~(\ref{eq:exactdiff}) also shows
that $P(x)$ is normalized to 1:
\be
\int d^d x\, P(x)
=\int_0^{+\infty}dp\int d^d x\, P(x)\,\delta(p(x)-p)
=\Delta(\infty)-\Delta(0)=1\;.
\ee
In principle, 
the uniform generation of events is therefore straightforward:
one generates a uniform random number $r$ between 0 and 1,
solves the equation $\Delta(p)=r$ for
$p$, and then generates values of $x$ on the surface
$\delta(p(x)-p)$ with a distribution proportional
to $R(x)\,\delta(p(x)-p)$.

In practice, however, the solution of the equation $\Delta(p)=r$
is in most cases very heavy from the numerical point of view.
This difficulty can be overcome by means of
the so-called vetoing method, which we now describe.
We assume that there is a function $H(x)\geq R(x)$ for all $x$ values,
and that
\begin{equation}
\Delta_H(p)=\exp\left[-\int d^dx^\prime\, H(x^\prime)\,
\theta(p(x^\prime)-p)\right]\;
\end{equation}
has a simple form, so that the solution of the equation
$\Delta_H(p)=r$ and the generation of the distribution $H(x)\,\delta(p(x)-p)$
are reasonably simple. Then, we implement the following procedure:
\begin{enumerate}
\item Set $p_{\rm max}=\infty$.
\item Generate a flat random number $0<n<1$.
\item Solve the equation
\begin{equation}
\frac{\Delta_H(p)}{\Delta_H(p_{\rm max})}=n
\end{equation}
for $p$
(a solution with $0<p<p_{\rm max}$ always exists for $0<n<1$).
\item Generate $x$ according to $H(x)\,\delta(p(x)-p)$.
\item Generate a new random number $n^\prime$.
\item
If $n^\prime>R(x)/H(x)$ then the event is vetoed, we
set $p_{\rm max}=p$, go to
step 3 and continue; otherwise the event is accepted, and the procedure stops.
\end{enumerate}
The resulting events are distributed according to
eq.~(\ref{eq:distr}).  The proof of this statement is simple but
non-trivial. At the end of the vetoing procedure, the event
distribution will be given by the sum of the distribution for the case
in which there is no veto, there is one veto applied, two vetoes, etc..
The probability distribution of events generated with no veto applied is
given by
\be
P_0(x)=\int_0^{p_{\rm max}}dp_1\,
\frac{\Delta_H(p_1)}{\Delta_H(p_{\rm max})}\,
H(x)\,\delta(p(x)-p_1))\,\frac{R(x)}{H(x)}
\nonumber\\
=R(x)\,\Delta_H(p(x))\,.
\ee
We have used the fact that $\Delta_H(p_{\rm max})=\Delta_H(\infty)=1$,
and we have inserted a factor of $R(x)/H(x)$, corresponding to the acceptance
probability.

When one veto is applied, we have
\ba
P_1(x)&=&\int_0^{p_{\rm max}}dp_1\,
\frac{\Delta_H(p_1)}{\Delta_H(p_{\rm max})}\,
\int d^dx_1\,H(x_1)\,\delta(p(x_1)-p_1)\,\left(1-\frac{R(x_1)}{H(x_1)}\right)\,
\nonumber\\
&&
\int_0^{p_1}dp_2\, \frac{\Delta_H(p_2)}{\Delta_H(p_1)}\,
H(x)\,\delta(p(x)-p_2))\,\frac{R(x)}{H(x)}
\nonumber\\
&=&R(x)\,\Delta_H(p(x))\,\int_{p(x)}^{+\infty}dp_1\,h(p_1)\,,
\label{eq:oneveto}
\ea
where we have defined
\begin{equation}
h(p_1)=\int d^dx_1\,H(x_1)\,
\delta(p(x_1)-p_1)\,\left(1-\frac{R(x_1)}{H(x_1)}\right)\,.
\end{equation}
The factor $1-R(x_1)/H(x_1)$ is the rejection probability,
which must be inserted at each vetoed step. Note that 
the result is nonzero only for $p_1\geq p(x)$, because of the
$\delta$ function in the $p_2$ integration.
It will be useful to perform one more step explicitly; for two vetoes,
we find
\ba
P_2(x)&=&\int_0^{p_{\rm max}}dp_1\,
\frac{\Delta_H(p_1)}{\Delta_H(p_{\rm max})}\,h(p_1)
\int_0^{p_1}dp_2\, \frac{\Delta_H(p_2)}{\Delta_H(p_1)}\,h(p_2)
\nonumber\\
&&
\int_0^{p_2}dp_3\,
\frac{\Delta_H(p_3)}{\Delta_H(p_2)}\,
H(x)\,\delta(p(x)-p_3))\,\frac{R(x)}{H(x)}
\nonumber\\
&=&R(x)\,\Delta_H(p(x))\,\frac{1}{2}
\left[\int_{p(x)}^{+\infty}dp\,h(p)\right]^2\,,
\label{eq:twoveto}
\ea
where we have used symmetric integration.
It is now easy to obtain the generic term of this infinite sum,
namely the term with $n$ vetoes applied. We get
\be
P_n(x)= \Delta_H(p(x))\,R(x)\,\frac{1}{n!}
\left[\int_{p(x)}^{+\infty}dp\,h(p)\right]^n\,.
\ee
The sum over $n$ yields
\ba
\sum_{n=0}^\infty P_n(x)&=&
R(x)\,\Delta_H(p(x))\,
\exp\left[\int_{p(x)}^{+\infty} dp\,h(p)\right]
\nonumber\\
&=&
R(x)\,\Delta_H(p(x))\,
\exp\left[\int d^d x^\prime\, [H(x^\prime)-R(x^\prime)]\,
\theta(p(x^\prime)-p(x))\right]
\nonumber\\
&=&
R(x)\,\exp\left[-\int d^d x^\prime\, R(x^\prime)\,
\theta(p(x^\prime)-p(x))\right]\,,
\ea
which is the announced result.

\section{Generation of events according to $\Delta_q^{(U)}$}
\label{app:Uq}
The veto technique described in the previous Appendix can be employed
to find the solution of eq.~(\ref{eq:genU}),
which we reproduce here: 
\begin{equation}\label{eq:genUapp}
n=\frac{\Delta_q^{(U)}(v,\pt)}{\Delta_q^{(U)}(v,p_{\rm max})}
\end{equation}
where $n$ is a random number between 0 and 1, 
\be
\Delta_q^{(U)}(v,\pt)=\exp\left[-\int U_q(v,r)\,
\theta(\kt(v,r)-\pt)\, d\Phi_r\right]\,;\qquad
r=\{x,y,\thd\}
\ee
and
\begin{equation}\label{eq:Maggbis}
U_q(v,r)=N_q \, \frac{\as(\kt^2(v,r))}{(1-x)\,(1-y^2)}\,.
\end{equation}
Trading the variable $y$ for
\be
\label{eq:kty}
\kt^2=\kt^2(v,r)=\frac{\mzz^2}{4x}\,(1-x)^2(1-y^2),\qquad
\abs{y}=\sqrt{1-\frac{4x}{(1-x)^2}\frac{\kt^2}{\mzz^2}}
\ee
we find
\ba
\int U_q(v,r)\,\theta(\kt(v,r)-\pt)\, d\Phi_r&=&
\int_\rho^1dx \int_{-1}^1 dy \int_0^\pi d\thd \,U_q(v,r)\,
\theta(\kt(v,r)-\pt)
\nonumber\\
&=&\pi N_q \,\int_\rho^{x_-}dx\,\int_{\pt^2}^{{\kt^2}_{\rm max}}
\frac{d\kt^2}{\kt^2}\,\frac{\as(\kt^2)}{\sqrt{(x_+-x)(x_--x)}}\,,
\ea
where 
\be
\rho=\frac{\mzz^2}{S};\qquad {\kt^2}_{\rm max}=\frac{S}{4}(1-\rho)^2;
\qquad
x_\pm=\left(\sqrt{1+\frac{\kt^2}{\mzz^2}}\pm\frac{\kt}{\mzz}\right)^2.
\ee
We have supplied an extra factor of 2 to account
for the fact that each value of $\kt^2$ corresponds to two values of $y$.
The integration range for $x$ is limited by
the definition $x=\mzz^2/s$, and by the condition $s\leq S$.
The $x$ integration can be performed
with the variable change $\xi=\sqrt{x_+-x}+\sqrt{x_--x}$.
We get
\be
\int U_q(v,r)\,\theta(\kt(v,r)-\pt)\, d\Phi_r
=\int_{\pt^2}^{{\kt^2}_{\rm max}}
\frac{d\kt^2}{\kt^2}\,V(\kt^2)\,,
\ee
where
\be
V(\kt^2)=\pi N_q\,\as(\kt^2)\,\log\frac
{\sqrt{x_+-\rho}+\sqrt{x_--\rho}}
{\sqrt{x_+-\rho}-\sqrt{x_--\rho}}\,.
\ee
We now observe that
\be
\log\frac
{\sqrt{x_+-\rho}+\sqrt{x_--\rho}}
{\sqrt{x_+-\rho}-\sqrt{x_--\rho}}
\leq
\log\frac
{\sqrt{x_+}+\sqrt{x_-}}{\sqrt{x_+}-\sqrt{x_-}}
=\frac{1}{2}\log\frac{\kt^2+\mzz^2}{\kt^2}
\leq
\frac{1}{2}\log\frac{q^2}{\kt^2}\,
\ee
where
\be
q^2={\kt^2}_{\rm max}+\mzz^2=\frac{S}{4}(1+\rho)^2.
\ee
Furthermore,
\be
\as(\kt^2)=\frac{\as(\kt^2)}{\az(\kt^2)}\,\az(\kt^2)
\leq \az(\kt^2)\,,
\ee
where $\az(\kt^2)$ is the leading-log expression of the running coupling,
\be
\az(\kt^2)=\frac{1}{\beta_0\,\log\frac{\kt^2}{\Lambda^2}}\,.
\ee
Hence
\be
V(\kt^2)\leq \tilde V(\kt^2)=\frac{\pi N_q}{2\beta_0}\,
\frac{1}{\log\frac{\kt^2}{\Lambda^2}}\,\log\log\frac{q^2}{\kt^2}\,.
\ee
The $\kt^2$ integral of $\tilde V$ can be performed analytically:
\be
\int_{\pt^2}^{{\kt^2}_{\rm max}}
\frac{d\kt^2}{\kt^2}\,\tilde V(\kt^2)=
\frac{\pi N_q}{2\beta_0}
\left[\log\frac{q^2}{\Lambda^2}
\log\frac{\log\frac{{\kt^2}_{\rm max}}{\Lambda^2}}
{\log\frac{\pt^2}{\Lambda^2}}
-\log\frac{{\kt^2}_{\rm max}}{\pt^2}
\right].
\ee
We now proceed as follows:
\begin{enumerate}
\item We set $p_{\rm max}={\kt}_{\rm max}$.

\item\label{step2}
We generate uniformly a random number $n$, $0\leq n\leq 1$, and we
solve numerically the equation
\be
n=\frac{\Delta^{(\tilde V)}(\pt)}{\Delta^{(\tilde V)}(p_{\rm max})};
\qquad
{\Delta^{(\tilde V)}(\pt)}=\exp\left[-\int_{\pt^2}^{{\kt^2}_{\rm max}}
\frac{d\kt^2}{\kt^2}\,\tilde V(\kt^2)\right]
\ee
for $\pt$.

\item
We generate a second random number $n'$ between 0 and 1;
if $n'<V(\pt^2)/\tilde V(\pt^2)$ we accept the event,
otherwise we set $p_{\rm max}=\pt$ and we return to step~\ref{step2}.
\end{enumerate}

\end{document}